\documentclass[lettersize,journal]{IEEEtran}

\IEEEoverridecommandlockouts
\usepackage{algorithm} 		
\usepackage{algpseudocode}
\usepackage{cite}
\usepackage{amsmath,amssymb,amsfonts}
\usepackage{float}
\usepackage{graphicx}
\usepackage{textcomp}
\usepackage{xcolor}
\usepackage{stfloats}
\usepackage{array}
\usepackage{multirow}
\usepackage{caption}
\usepackage{subcaption}
\usepackage{flushend}
\usepackage{mathrsfs}
\usepackage{booktabs}
\usepackage{tabularx}

\usepackage{url}
\usepackage{verbatim}
\begin{document}
\title{Non-Orthogonal Affine Frequency Division Multiplexing for Spectrally Efficient High-Mobility Communications}

\author{
Qin Yi,~\IEEEmembership{Graduate~Student~Member,~IEEE,} Zilong Liu,~\IEEEmembership{Senior~Member,~IEEE,} Leila Musavian,~\IEEEmembership{Member,~IEEE,} and Zeping Sui,~\IEEEmembership{Member,~IEEE}
\thanks{Qin Yi, Zilong Liu, Leila Musavian, and Zeping Sui are with the School of Computer Science and Electronics Engineering, University of Essex, Colchester CO4 3SQ, U.K. (e-mail: qinyi17@outlook.com, {zilong.liu, leila.musavian}@essex.ac.uk, zepingsui@outlook.com)}
}

\maketitle
\begin{abstract}
This paper proposes a novel non-orthogonal affine frequency division multiplexing (nAFDM) waveform for reliable high-mobility communications with enhanced spectral efficiency (SE). The key idea is to introduce a bandwidth compression factor into the AFDM modulator to enable controllable subcarrier overlapping.
We first detail the proposed nAFDM transceiver and derive the corresponding input-output signal relationship.
Then, an efficient nAFDM signal generation method based on the inverse discrete Fourier transform (IDFT) is proposed, enabling practical implementation using existing inverse fast Fourier transform (IFFT) modules without additional hardware complexity. Next, to characterize the impact of non-orthogonal modulation, we derive a closed-form expression of inter-carrier interference (ICI), showing its dependence on the bandwidth compression factor. To mitigate the resulting interference, we propose a soft iterative detection algorithm and a low-complexity implementation approach that leverages the distribution characteristics of ICI. Simulation results demonstrate that 1) in terms of bit error rate (BER), the proposed nAFDM can achieve near identical BER compared to conventional AFDM, while outperforms other waveform counterparts; 2) nAFDM is capable of striking higher SE compared to other existing waveforms; and 3) the proposed nAFDM achieves an attractive BER vs. SE trade-off, and the proposed soft  {iterative detection (ID)} scheme can attain a trade-off between BER and complexity.
\end{abstract}

\section{Introduction}
\subsection{Background}
Providing reliable wireless connectivity in high-mobility scenarios, such as vehicle-to-everything systems, high-speed railways, unmanned aerial vehicles, and low-earth-orbit satellite networks, is crucial for the upcoming sixth generation systems \cite{HighMob1,HighMob2,HighMob3}. The widely adopted orthogonal frequency division multiplexing (OFDM) systems may suffer from severe inter-carrier interference (ICI) in high-mobility environments due to large Doppler {shifts}, leading to significant degradation in communication performance \cite{OFDM_Disadvan,OFDM_Disadvan_R1Add_1,OFDM_Disadvan_R1Add_2}. 
 {Against this problem, a two-dimensional modulation waveform called orthogonal time frequency space (OTFS) has been proposed to improve reliability in high-mobility channels, where information symbols are first  multiplexed onto the delay-Doppler grids and then transformed to the time-frequency domain via the inverse symplectic finite Fourier transform (ISFFT) \cite{OTFS_R1Add}. 
In addition, OTFS has been extensively investigated and further extended by integrating advanced transmission paradigms to improve spectral efficiency (SE), such as multiple-input multiple-output (MIMO) designs \cite{OTFS-MIMO_R1Add}, spatial modulation-aided schemes \cite{OTFS-SM_R1Add}, and OTFS-based multiple access frameworks \cite{OTFS-MA_R1Add}.
Unlike OTFS which involves two-dimensional delay-Doppler modulation, affine frequency division multiplexing (AFDM) emerges recently as a promising backwards-compatible  multicarrier waveform in \cite{AFDM}.}
In AFDM, data symbols are modulated over multiple orthogonal chirp subcarriers, thus spreading each information symbol across the entire bandwidth \cite{AFDM_R1Add_1}.
Such a modulation is carried out  {via} inverse discrete affine Fourier transform (IDAFT) \cite{DAFT1}, which can be efficiently implemented with the aid of inverse fast Fourier transform (IFFT) \cite{AFDM}. At the receiver side, DAFT is applied to recover the transmitted symbols. Since DAFT can be regarded as a generalization of both the discrete Fourier transform (DFT) and the discrete Fresnel transform (DFnT) \cite{DAFT_Special}, AFDM can be built upon the minimum modification of OFDM systems and includes orthogonal chirp division multiplexing (OCDM) as a special case \cite{OCDM}. 
 {Moreover, AFDM can be cast as a DFT-based precoded-OFDM waveform \cite{DFT-AFDM_R1Add_1} and was incorporated into a unified single-carrier interleaved frequency-division multiplexing (SC-IFDM)-based transceiver, enabling orthogonal coexistence and backwards-compatible implementations within existing FFT/IFFT modules \cite{DFT-AFDM_R1Add_2}.}
Compared to OFDM and OCDM, AFDM enables separable channel representation and achieves full diversity in doubly selective channels by appropriately tuning the chirp rate to match the Doppler profile \cite{AFDM,AFDM_Para1}.
These advantages make AFDM an attractive waveform  {for achieving} enhanced communication performance in high-mobility environments.

\subsection{Related Works}
Several detection and channel estimation algorithms have been proposed to enhance the performance of AFDM. These include a {maximum ratio combining} scheme \cite{AFDM,ZPAFDM}, an iterative minimum mean square error (MMSE) receiver that balances complexity and detection reliability \cite{AFDM_Para1}, and a {message passing} algorithm that exploits the sparse structure of the {DAFT-domain} channel to jointly perform interference cancellation and signal detection with reduced complexity \cite{AFDM_Detect3}. 
{Pilot}-aided channel estimation for AFDM systems was first introduced in \cite{AFDM}, where pilot symbols are embedded into the transmitted frame to efficiently estimate {channel parameters}. This approach was later extended to  {MIMO}-AFDM systems in \cite{AFDM_ChanEst1}, where a low-complexity diagonal reconstruction algorithm was developed to estimate the effective channel matrix. To enhance  {SE}, a channel estimation scheme leveraging superimposed pilots in the DAFT domain  {was} proposed \cite{AFDM_ChanEst2}.

Besides, AFDM has been integrated with other advanced transmission techniques for a wide range of applications. For example, it  {was} introduced to serve as the foundational multicarrier waveform to support massive high-mobility machine-type communications through the integration of {sparse code multiple access \cite{AFDM_SCMA}. In addition, AFDM with index modulation \cite{AFDM_IM1,AFDM_IM2} and generalized spatial modulation aided AFDM \cite{AFDM_GSM}}  {were} proposed for improved bit error rate (BER) performance and energy efficiency. 
 {Furthermore, AFDM has also been investigated in the context of integrated sensing and communications (ISAC). 
In \cite{AFDM_ISAC1}, an AFDM-based ISAC system was developed and it was shown that accurate range-velocity sensing can be achieved using only a single DAFT-domain pilot. 
The ambiguity function properties of AFDM signals for range-velocity sensing were later analyzed in \cite{AFDM_ISAC_R1Add2-01}, and two performance metrics for AFDM-ISAC, namely sensing SE and sensing outage probability, were introduced in \cite{AFDM_ISAC2}. 
In \cite{AFDM_ISAC3}, a Bayesian parametric bilinear Gaussian belief propagation framework was proposed for AFDM-ISAC to enable joint channel estimation, data detection, and radar-parameter estimation.
In addition, recent studies have explored physical-layer security for AFDM transmissions by exploiting the flexibility of chirp parameters, including secret pre-chirp sequence permutation \cite{AFDM_PLS_R1Add1} and reciprocity-based user-specific pre-chirp parameter generation \cite{AFDM_PLS_R1Add2}.
Several modified affine-waveform variants have been proposed to improve AFDM characteristics. 
In particular, affine filter bank modulation was proposed in \cite{AFBM_R1Add1,AFBM_R1Add2}, which integrates filter bank multicarrier processing into the AFDM framework to reduce the peak-to-average power ratio and out-of-band emission.
Chirp-permuted AFDM was introduced in \cite{Chirp-permutedAFDM_R1Add1} by incorporating a chirp-permutation domain on top of conventional AFDM, thereby enhancing ambiguity function resolution and the peak-to-sidelobe ratio in the Doppler domain.}

To meet the ever-growing demand for higher SE, non-orthogonal multicarrier waveform designs have attracted increasing research attention. One notable example is spectrally efficient frequency division multiplexing (SEFDM), which enhances spectral utilization by employing non-orthogonal overlapped subcarriers in the frequency domain \cite{SEFDM1,SEFDM2}. This enables the transmission of more symbols within a given bandwidth compared to conventional OFDM systems. To support practical implementation, an efficient inverse DFT (IDFT)-based method  {was} proposed in \cite{SEFDM_SignalGen} to generate SEFDM signals with complexity similar to that of OFDM. 
However, the orthogonality loss in SEFDM introduces inter-carrier interference (ICI), which  {was} characterized through mathematical analysis and simulation modeling in \cite{SEFDM_ICI}.
Linear detectors such as zero forcing \cite{SEFDM_ZF} and MMSE \cite{SEFDM_MMSE_SVD}  {were} investigated for their simplicity, but suffer from degraded error rate performance under strong ICI. To improve  {the performance of the MMSE detector}, an iterative detection (ID) method  {was} proposed in \cite{SEFDM_ID} to progressively mitigate interference. Nonetheless, as a non-orthogonal extension of OFDM, SEFDM inherits sensitivity to Doppler effects, leading to notable performance degradation under rapid channel variations and limiting its applicability in real-time high-mobility scenarios.

\subsection{Motivations and Contributions}
Although AFDM has demonstrated strong robustness to doubly selective channel variations, most existing studies have focused on orthogonal chirp subcarriers. To attain a higher spectrum efficiency, it is intriguing to understand 1) how non-orthogonal chirp subcarriers can bring in further performance gain? and 2) what are the effective techniques to {mitigate} the interference incurred by non-orthogonality? To address these two questions, we propose a novel non-orthogonal AFDM (nAFDM) waveform that combines {the improved} SE benefits of non-orthogonal signaling with the inherent resilience {to high-mobility scenarios}. Unlike SEFDM, nAFDM operates in the affine time-frequency domain.
The main contributions of this work are summarized as follows:
\begin{itemize}
\item We present a generic transceiver framework for the proposed nAFDM waveform. By introducing a bandwidth compression factor in the affine time-frequency domain, the proposed nAFDM enhances SE through controllable subcarrier overlapping, while maintaining robustness in doubly selective channels.
Moreover, the corresponding input-output relationship is derived to characterize the impact of the bandwidth compression factor on channel characteristics, revealing the distinctive energy spreading induced by non-orthogonal transmission.

\item We introduce an efficient signal generation method based on IDFT for nAFDM and show that it achieves implementation complexity comparable to that of conventional AFDM systems without additional hardware overhead. The corresponding demodulation is implemented via DFT, and both operations can be efficiently realized using existing IFFT/FFT modules to enable rapid processing. To quantify the non-orthogonality induced interference, we derive a closed-form expression {of} the resulting ICI and demonstrate its dependence on the bandwidth compression factor, providing insights into the ICI behavior introduced by non-orthogonal modulation.

\item Building {upon} the ICI analysis, we develop a soft ID algorithm that incorporates symbol probability information into the interference cancellation process, enabling progressive refinement of symbol estimates and effective ICI suppression. A redetection step is further introduced to {refine the decisions of those} low-reliability symbols based on their estimated variances and soft probabilities, thereby improving detection accuracy. 
{Then}, we propose a low-complexity ICI pruning technique that exploits the distribution of {those dominant interference components}, thereby achieving a favorable trade-off between performance and complexity.

\item Finally, extensive simulation results are presented to demonstrate the advantages of the proposed nAFDM system. Specifically, we show that 1) nAFDM significantly outperforms SEFDM and conventional orthogonal schemes such as OFDM and OCDM {with regard to} both BER and SE under doubly selective channels; 
2) It achieves BER performance close to that of {conventional} AFDM while providing substantial SE gains; 
3) The proposed soft ID approach notably improves BER performance compared to existing MMSE and ID methods; 
and 4) {We} provide a comprehensive BER analysis under various bandwidth compression factors, offering practical insights on {the} selection of suitable compression parameters to achieve an improved trade-off between BER and SE.


\end {itemize}

\subsection{Organization}
The rest of this paper is organized as follows: Section~\ref{Sec:SystemModel} presents the system model of the proposed nAFDM scheme. Section~\ref{Sec:SignalGen} describes the DFT-based signal generation algorithm and discusses its implementation complexity.
In Section~\ref{Sec:ICIAnalysis_Dec}, we analyze the ICI introduced by non-orthogonal modulation and develop the proposed soft ID algorithm based on this analysis. Simulation results are presented in Section~\ref{Sec:Results}, and conclusions are drawn in Section~\ref{Sec:Conclusion}.

\subsection{Notation} 
Bold lowercase and uppercase letters denote vectors and matrices, respectively. $\mathbb{C}^{M \times N}$ represents the space of complex-valued matrices of size $M \times N$. $\operatorname{diag}(\mathbf{a})$ denotes a diagonal matrix with the elements of vector $\mathbf{a}$ on its main diagonal. The operators $(\cdot)^*$, $(\cdot)^T$, $(\cdot)^H$, and $(\cdot)^{-1}$ denote the complex conjugate, transpose, Hermitian, 
and matrix inverse, respectively. ${\bf{I}}_M$ denotes the $M \times M$ identity matrix. {$\delta(\cdot)$ represents the Dirac delta function, and $(\cdot)_N$ denotes the modulo-$N$ operation.}  {$\Re\{\cdot\}$ and $\Im\{\cdot\}$} denote the real and imaginary parts of a complex quantity, respectively. The expectation operator is denoted by $\mathbb{E}\{\cdot\}$. 

\captionsetup[figure]{singlelinecheck=off}
\begin{figure*}
    \centering
    \includegraphics[width=0.95\linewidth]{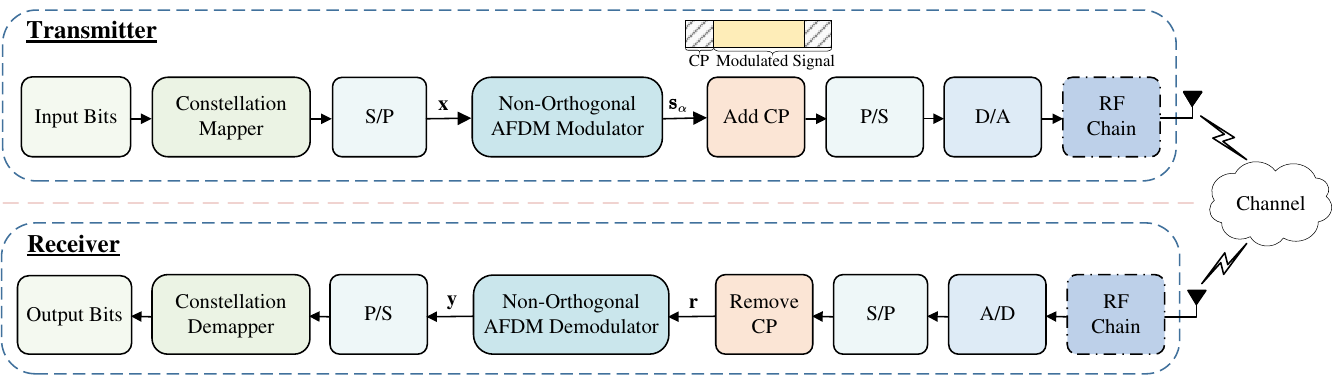}
    \captionsetup{subrefformat=parens} 
    \caption[left]{System block diagram of the proposed nAFDM waveform.}
    \label{fig:SystemModel}
\end{figure*}
\captionsetup[figure]{singlelinecheck=on}

\section{System Model}\label{Sec:SystemModel}
In this section, we first describe the modulation and demodulation procedures of the proposed nAFDM waveform, where a bandwidth compression factor is introduced to enable subcarrier overlap and improve SE. To address the loss of chirp periodicity due to non-orthogonality, a cyclic prefix (CP) is adopted to mitigate multipath interference. We then derive the corresponding input-output relationship, which reveals a unique energy spreading behavior induced by the combined effects of delay, Doppler shift, and bandwidth compression factor.

\subsection{Modulation}
The transceiver structure of the proposed nAFDM system is illustrated in Fig.~\ref{fig:SystemModel}. At the transmitter, the information symbol vector $\mathbf{x} \in \mathbb{C}^{N \times 1}$ in the affine frequency domain is mapped to a set of non-orthogonal chirp subcarriers in the time domain by non-orthogonal AFDM modulation, which is expressed as
 \begin{align}
{\label{eq:Modu_Sym}
    {s}_{\alpha}[n]=\frac{1}{\sqrt{N}}\sum\limits_{m=0}^{N-1} {x}[m] e^{\imath2\pi\left(c_1n^2+c_2m^2+\frac{\alpha nm}{N}\right)}, }
\end{align} 
{for $n=0, \ldots, N-1$,} where \(s_{\alpha}[n]\) denotes the data symbol modulated on the \(n\)th chirp subcarrier, $N$ is the number of chirp subcarriers, {while} $c_1$, $c_2$ are the chirp parameters controlling the quadratic phase rotation in the time and frequency domains, respectively. 
{Moreover,} $\alpha$ represents the bandwidth compression factor, defined as $\alpha = \Delta f T$, where \(\Delta f\) is the subcarrier spacing and \(T\) denotes the symbol duration. It is noted that conventional orthogonal AFDM waveform corresponds to $\alpha = 1$ \cite{AFDM}, whilst \(\alpha\in(0,1)\) {leads} to nAFDM.

The time-domain signal in \eqref{eq:Modu_Sym} can be equivalently represented in matrix form as
\begin{align}\label{eq:TxSigal}
    \mathbf{s}_{\alpha} = \mathbf{A}_\alpha^H\mathbf{x}=\mathbf{\Lambda}_{c_1}^H \mathbf{F}_\alpha^H \mathbf{\Lambda}_{c_2}^H \mathbf{x},
\end{align}
where {$\mathbf{A}_{\alpha} = \mathbf{\Lambda}_{c_2} \mathbf{F}_\alpha \mathbf{\Lambda}_{c_1}$} represents the modulation matrix for nAFDM, with  $\mathbf{\Lambda}_{c} = \text{diag}\left(e^{-\imath 2\pi c n^2}, \, n = 0, \ldots, N-1 \right)$ and $\mathbf{F}_\alpha$ having {entries of} $\frac{1}{\sqrt{N}} e^{-\imath 2\pi \frac{\alpha mn}{N}}$.  {Moreover, $\mathbf{F}_\alpha$ can be viewed as the scaled principal $N\times N$ submatrix of an $N'$-point DFT matrix $\bar{\mathbf{F}}\in\mathbb{C}^{N'\times N'}$ with $N'=N/\alpha$, see \eqref{eq:DFT} in Appendix~B.}


\subsection{CP}
In conventional orthogonal AFDM systems, the time-domain signal exhibits inherent chirp-periodicity, which enables the application of a chirp-periodic prefix (CPP) to combat multipath propagation \cite{AFDM}. However, in the nAFDM system, the time-domain signal satisfies
\begin{align}\label{eq:s_nPlusKN}
s_{\alpha}[n + \tau N] &= \frac{1}{\sqrt{N}} e^{\imath2\pi c_1(\tau^2N^2 + 2\tau Nn)}  \nonumber \\
&\times \sum_{m=0}^{N-1} e^{\imath2\pi\alpha m\tau} \cdot x[m] e^{\imath2\pi\left(c_1 n^2 + c_2 m^2 + \frac{\alpha n m}{N} \right)},
\end{align}
where \(\tau \in \mathbb{Z}\) denotes the period index. Due to the presence of the term $e^{\imath 2\pi \alpha m\tau} \ne 1$ when \(\alpha\in(0,1)\), \eqref{eq:s_nPlusKN} {cannot} be simplified to a phase-shifted version of $s_\alpha[n]$, i.e., $s_{\alpha}[n + \tau N] \ne e^{\imath2\pi c_1(\tau^2N^2 + 2\tau Nn)} s_\alpha[n]$, which indicates that the chirp-periodic structure is no longer preserved, thereby rendering the CPP approach {is} incompatible. To address this issue, the proposed nAFDM adopts a conventional CP instead of a CPP, which is defined as
\begin{align}\label{eq:CP}
s_\alpha[n] = s_\alpha[N + n],n=-L_{\mathrm{CP}},\ldots,-1,
\end{align}
where \(L_{\mathrm{CP}}\) denotes the CP length, which is required to be at least {the  maximum delay spread (in samples) of the channel} to ensure effective ISI mitigation.

\subsection{Channel Model}
The channel impulse response of the $P$-path doubly selective channel  {at time $n$ and delay $l$} is expressed as
\begin{align}\label{eq:channel}
g_n(l) = \sum_{i=1}^{P} h_i e^{-\imath {2\pi} f_i n} \delta(l-l_i),
\end{align}
{where $h_i$, $l_i$, and $f_i$ denote the channel fading coefficients,}  integer delay shifts, and Doppler shifts (in digital frequencies) of the $i$-th path, respectively. Let \(\nu_i \triangleq Nf_i = a_i + \beta_i\) represent the Doppler shift normalized with respect to the subcarrier spacing, 
 {where $a_i\in\mathbb{Z}$ is the integer part of  $\nu_i$ with $a_i \in [-a_{\max}, a_{\max}]$, and  \(\beta_i \in \left(-\frac{1}{2}, \frac{1}{2}\right]\) denotes the fractional part.}
Note that $\nu_{\max}\triangleq{\text{max}}(\nu_i)$ and $l_{\max}\triangleq{\text{max}}(l_i)$ represent the normalized maximum Doppler shift and the maximum delay, respectively.

\subsection{Demodulation}
At the receiver, the time-domain  received signal can be expressed as
\begin{align}\label{eq:RxSym_TD}
    r[n] = \sum_{l=0}^{\infty} s_\alpha[n-l] g_n(l) + w_\mathrm{T}[n],
\end{align}
where $n_\mathrm{T}\sim \mathcal{CN}(0, \sigma^2)$ represents the additive white Gaussian noise (AWGN), {and $\sigma^2$ denotes the noise variance}. After removing CP, \eqref{eq:RxSym_TD} can be rewritten in the matrix form as
\begin{align}\label{eq:RxSignal_TD}
{\mathbf{r}}  = {\mathbf{H}}_\mathrm{T} {\mathbf{s}}_\alpha + {\mathbf{w}}_\mathrm{T},
\end{align} 
{where the time-domain channel matrix can be given by}
\begin{align}\label{eq:Channel_TD}
{\mathbf{H}}_\mathrm{T} = \sum_{i=1}^{P} h_i {\mathbf{\Delta}}_{f_i} \boldsymbol{\Pi}^{l_i},
\end{align} 
with ${\mathbf{\Delta}}_{f_i} \triangleq \operatorname{diag}\left(e^{-\imath 2 \pi f_i n}, n = 0, 1, \ldots, N-1\right)$,  and $\boldsymbol{\Pi}$ is the $N \times N$ forward cyclic-shift matrix

\begin{align}
\label{eq:Chan_ForCycShiftMat}
\boldsymbol{\Pi} =
\begin{bmatrix}
0 & \cdots & 0 & 1 \\
1 & \cdots & 0 & 0 \\
\vdots & \ddots & \vdots & \vdots \\
0 & \cdots & 1 & 0
\end{bmatrix}.
\end{align} 
The time-domain received signal vector is then transformed into the affine frequency domain as
 \begin{align}\label{eq:RxSignal_AFD}
{\mathbf{y}} = {\mathbf A}_\alpha{\mathbf{r}}  ={\mathbf H}_\mathrm{eff} {{\bf{x}}} + {\mathbf{w}}_\mathrm{AF},
\end{align}
 where ${\mathbf H}_\mathrm{eff}={\mathbf A}_\alpha{\mathbf{H}}_\mathrm{T} {\mathbf A}_\alpha^H$ denotes the effective {channel matrix. Moreover,} ${\mathbf{w}}_\mathrm{AF} = {\mathbf A}_\alpha{\mathbf{w}}_\mathrm{T}$ represents the transformed noise in the affine frequency domain, whose correlation matrix ${\mathbf R}_w = \mathbb{E}\{{\mathbf{w}}_\mathrm{AF}{\mathbf{w}}_\mathrm{AF}^H\}=\sigma^2{\mathbf{A}}_\alpha{\mathbf{A}}_\alpha ^H$ is generally non‑diagonal due to the non-unitary nature of $\mathbf{A}_\alpha$. Consequently, the noise becomes colored, exhibiting correlation across components, in contrast to the uncorrelated nature of conventional white noise.
 

\begin{figure*}[!htbp]
    \centering
    \begin{subfigure}{0.168\textwidth} 
        \centering
        \includegraphics[width=\linewidth]{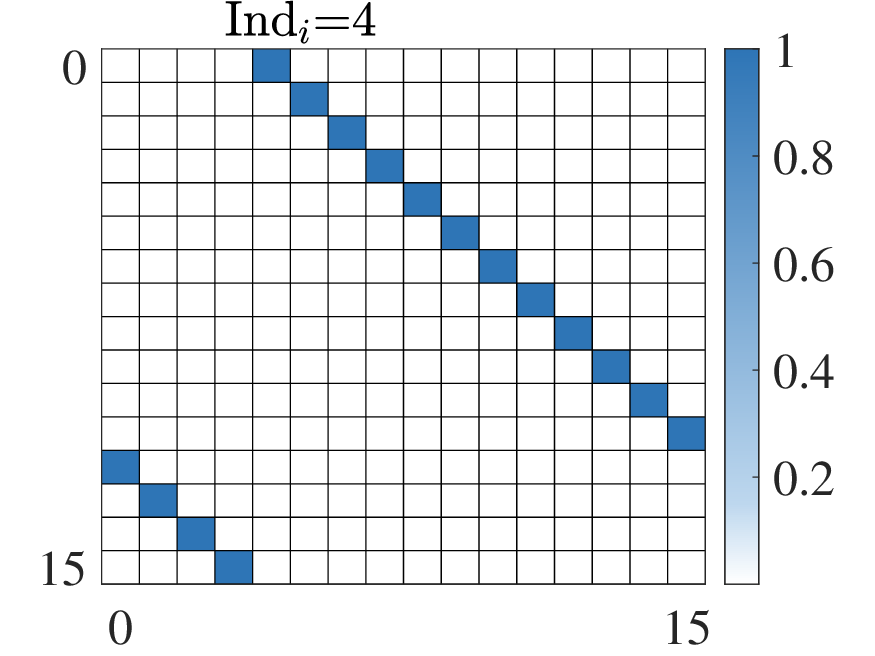}
        \subcaption{$\nu_i=1$}
        \label{fig:AfdmHeffIntDopp}
    \end{subfigure}
    \hspace*{-1.1em} 
    \begin{subfigure}{0.168\textwidth}
        \centering
        \includegraphics[width=\linewidth]{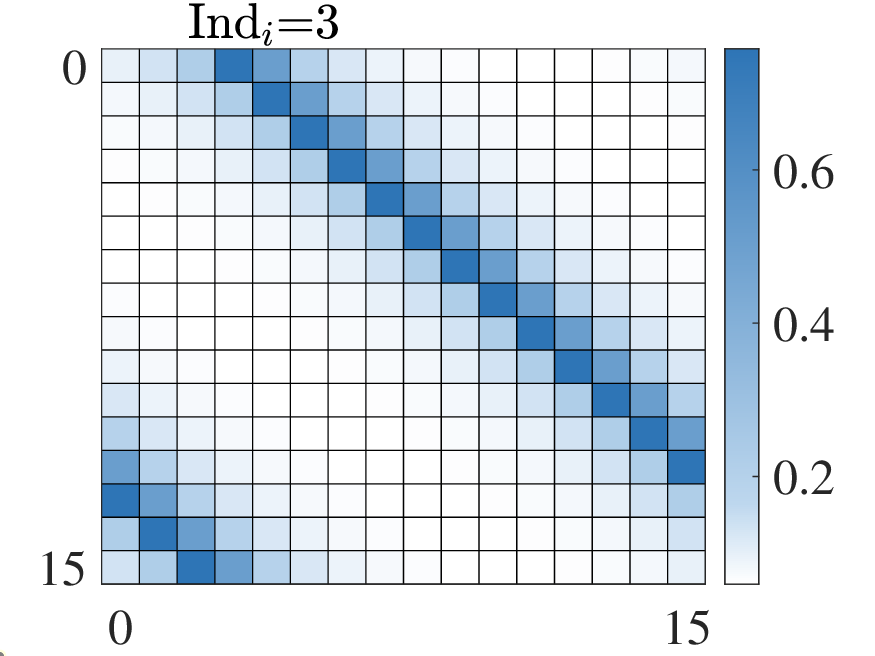}
        \subcaption{$\nu_i=0.4$}
        \label{fig:AfdmHeffFracDop}
    \end{subfigure}
    \hspace*{-1.1em}
    \begin{subfigure}{0.168\textwidth} 
        \centering
        \includegraphics[width=\linewidth]{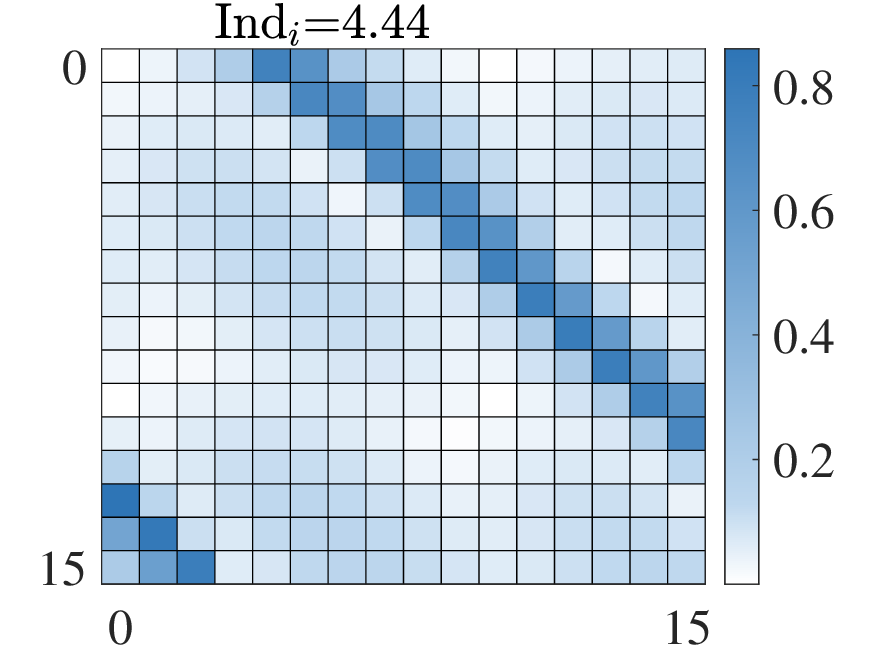}
        \subcaption{$\nu_i=1$}
        \label{fig:NonAfdmHeffIntDopp90}
    \end{subfigure}
    \hspace*{-1.1em}
    \begin{subfigure}{0.168\textwidth}
        \centering
        \includegraphics[width=\linewidth]{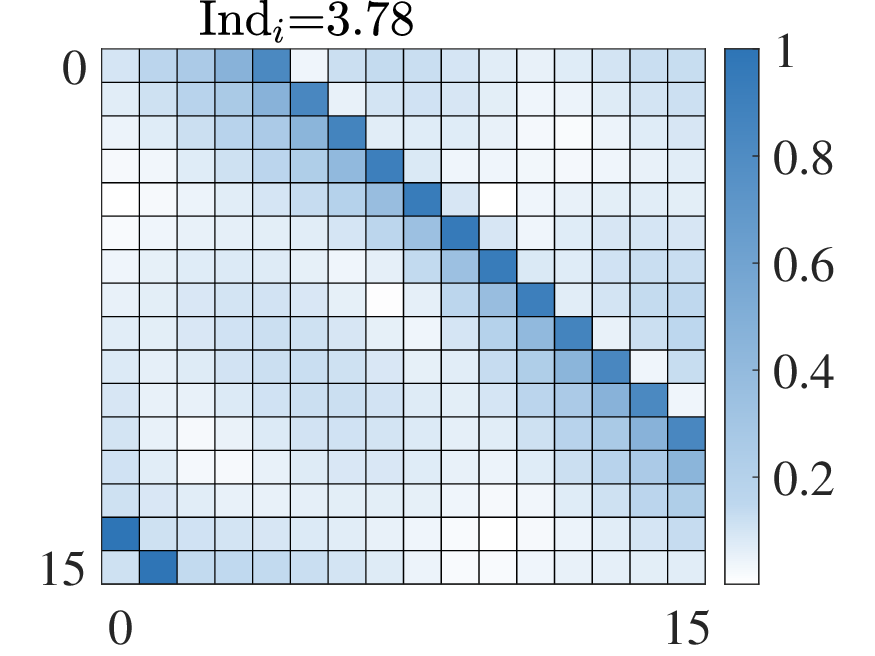}
        \subcaption{$\nu_i=0.4$}
        \label{fig:NonAfdmHeffFracDopp90}
    \end{subfigure}
        \hspace*{-1.1em}
    \begin{subfigure}{0.168\textwidth} 
        \centering
        \includegraphics[width=\linewidth]{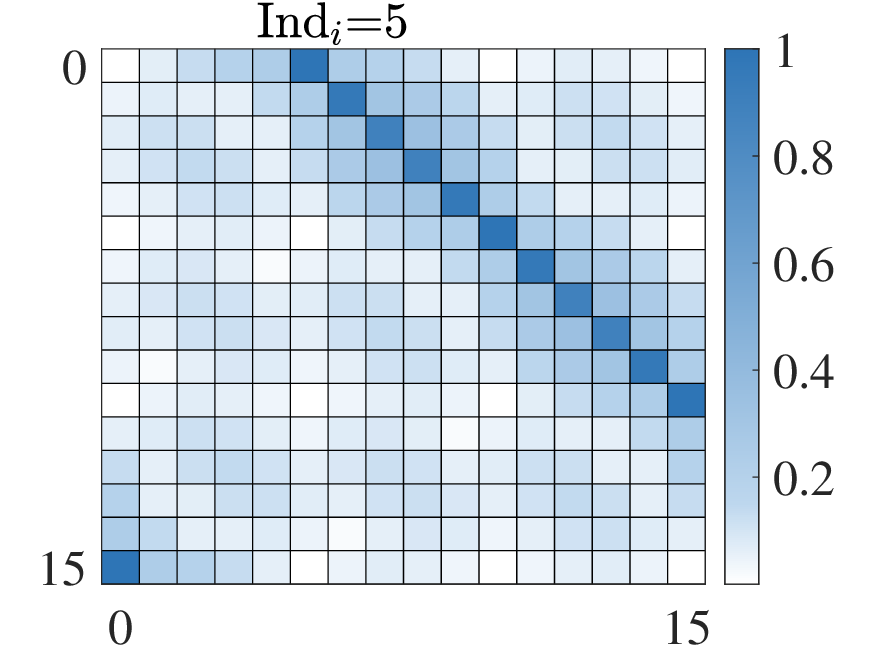}
        \subcaption{$\nu_i=1$}
        \label{fig:NonAfdmHeffIntDopp80}
    \end{subfigure}
    \hspace*{-1.1em}
    \begin{subfigure}{0.168\textwidth}
        \centering
        \includegraphics[width=\linewidth]{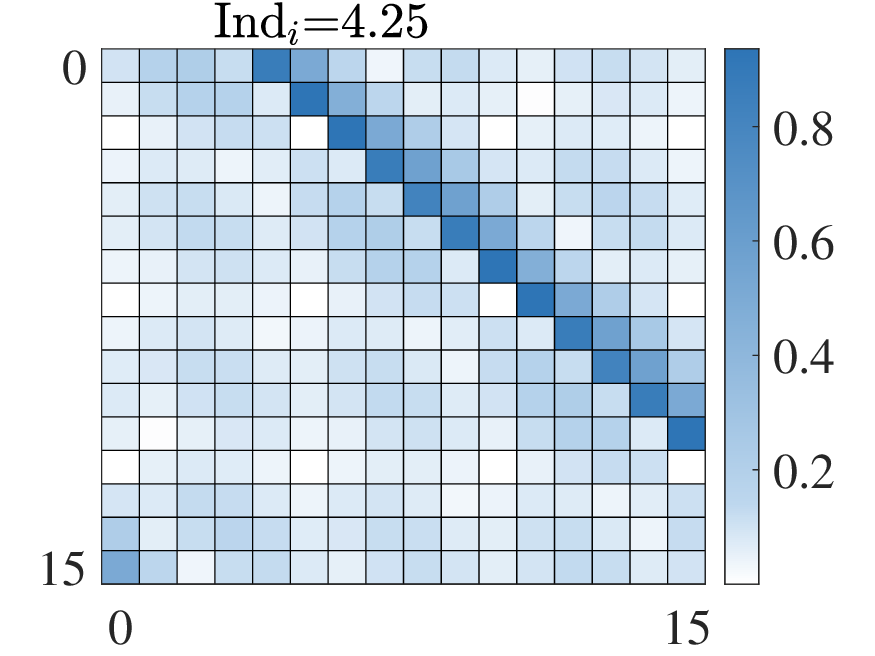}
        \subcaption{$\nu_i=0.4$}
        \label{fig:NonAfdmHeffFracDopp80}
    \end{subfigure}   
    \captionsetup{subrefformat=parens} %
    \caption{Structure of the effective channel matrix $\mathbf{H}_i$ under different Doppler shifts and compression factors: 
(a) AFDM with $\nu_i=1$; 
(b) AFDM with $\nu_i=0.4$; 
(c) nAFDM ($\alpha=0.9$) with $\nu_i=1$; 
(d) nAFDM ($\alpha=0.9$) with $\nu_i=0.4$; 
(e) nAFDM ($\alpha=0.8$) with $\nu_i=1$; 
(f) nAFDM ($\alpha=0.8$) with $\nu_i=0.4$. 
The system parameters are set to $N=16$, $l_i=1$, and $c_1=c_2=\frac{3}{2N}$.}
\label{fig:H_eff}
\end{figure*}

\subsection{Input-Output Relationship}
Substituting \eqref{eq:TxSigal}, \eqref{eq:RxSignal_TD}, and \eqref{eq:Channel_TD} into \eqref{eq:RxSignal_AFD}, the received signal in the affine frequency domain is given by
\begin{align}\label{eq:RxSignal_AFD2}
{\mathbf{y}}  =\sum_{i=1}^{P} h_i {\mathbf{H}}_i \bf{x} + {\mathbf{w}}_\mathrm{AF},
\end{align}
where ${\mathbf{H}}_i={\mathbf A}_\alpha  {\mathbf\Delta}_{f_i} \boldsymbol{\Pi}^{l_i}{\mathbf A}_\alpha^H$ represents the subchannel matrix of the $i$-th path. The $(p, q)$ entry of ${\mathbf{H}}_i$ is expressed as
\begin{align}\label{eq:SubHeff_Hi}
H_i[p,q]
=\frac{1}{N}
\eta(\alpha,l_i,p,q)\zeta (\alpha,l_i,\nu_i,p,q),
\end{align}
with
\begin{align}\label{SubHeff_Hi_Part1}
\eta(\alpha,l_i,p,q)
=
e^{{\imath}\frac{2\pi}{N}\left [
Nc_1l_i^2-\alpha\,q\,l_i
+N c_2(q^2-p^2)\right ]},
\end{align}
\begin{align}\label{SubHeff_Hi_Part2}
\zeta (\alpha,l_i,\nu_i,p,q)
&=(e^{\imath2\pi\alpha q}-1)\sum_{n=0}^{l_i-1}
  e^{-\imath\frac{2\pi}{N}\phi   n}+\sum_{n=0}^{N-1}
  e^{-\imath\frac{2\pi}{N}\phi   n}\nonumber\\
&=\frac{(e^{\imath2\pi\alpha q}-1)(e^{-\imath\frac{2\pi}{N} l_i\phi  }-1)+ (e^{-\imath{2\pi}\phi  }-1)}
{e^{-\imath\frac{2\pi}{N}\phi  }-1},
\end{align}
where $\phi  ={\alpha(p-q)+\nu_i+2Nc_1l_i}$.

{\emph{Proof:} The detailed derivation of ${\mathbf{H}}_i$ is provided in \textbf{Appendix A}.} \hfill $\blacksquare$

Moreover, $\eta(\alpha,l_i,p,q)$ contributes only to phase (without altering the amplitude), while the compression factor $\alpha$ affects the amplitude response through \eqref{SubHeff_Hi_Part2},  thereby influencing the energy distribution across subchannels. Thus, the input-output {relationship} for nAFDM is written as
\begin{align}
y[p] =& \sum_{\substack{q=0}}^{N-1} 
        \sum_{\substack{i=1}}^{P} 
        \frac{h_{i}}{N} \eta(\alpha,l_i,p,q) \zeta (\alpha,l_i,\nu_i,p,q) x[q] + w_{\mathrm{AF}}[p].
\end{align}
In particular, when \(\alpha=1\), we have \(e^{\imath2\pi\alpha q}=1\), and \eqref{SubHeff_Hi_Part1} and \eqref{SubHeff_Hi_Part2} simplify to 
\begin{align}
\eta(l_i,p,q)=e^{{\imath}\frac{2\pi}{N}\left [Nc_1l_i^2-q\,l_i+N c_2(q^2-p^2)\right ]}, 
\end{align}
\begin{align}
\zeta (l_i,\nu_i,p,q)=\sum_{n=0}^{N-1}
  e^{-\imath\frac{2\pi}{N}\phi   n}=\frac{e^{-\imath2\pi \phi  }-1}{e^{-\imath\frac{2\pi}{N}\phi  }-1},
\end{align}
respectively, which correspond to the conventional orthogonal AFDM subchannel response. Based on \eqref{eq:SubHeff_Hi} - \eqref{SubHeff_Hi_Part2}, the magnitude of $H_i[p,q]$ can be expressed as
\begin{align}
|H_i[p,q]| = \frac{1}{N} |\zeta (\alpha,l_i,\nu_i,p,q)|.
\end{align}
Thus, the magnitude of $H_i[p,q]$ reaches its peak at $q^* = (p + \mathrm{Ind}_i)_{(N/\alpha)}$ and gradually decreases as $q$ moves away from $(p + \mathrm{Ind}_i)_{(N/\alpha)}$, where $\mathrm{Ind}_i\triangleq \left [{(\nu_i + 2N c_1 l_i)/{\alpha}} \right ]_{(N/\alpha)}$. For nAFDM, since \(\alpha\in(0,1)\), the ideal peak location $q^*$ may not correspond to an integer.  However, \(q\) must be an integer column index, and hence the actual maximum magnitude will occur at the integer \(q\) closest to \( q^*\). In the special case where $\alpha = 1$, $|H_i[p,q]|$ reaches its peak when $q = (p + \mathrm{Ind}_i)_N$ with $\mathrm{Ind}_i=(a_i + 2N c_1 l_i)_N$.

Fig.~\ref{fig:H_eff} shows the structure of the channel matrix \( \mathbf{H}_i \) under different Doppler conditions and compression factors. For conventional AFDM, when Doppler shifts are integers, each row or column of $\mathbf{H}_i$ contains exactly one non-zero entry. On the other hand, for non-integer Doppler shifts, a small cluster of secondary non-zero entries appears adjacent to the main path due to fractional Doppler leakage. 
In contrast, in nAFDM, due to the non-integer nature of \(\alpha\), it becomes difficult for \(p\) and \(q\) to precisely satisfy the integer alignment condition, causing the main path energy to split and spread across multiple adjacent positions.
 {Furthermore, the distribution pattern of \(\mathbf{H}_i\) under integer Doppler shifts exhibits a spreading behavior similar to that in the fractional Doppler case, as the combined effect of the Doppler shift and the bandwidth compression factor introduces fractional phase components even for integer Doppler values.}
As a result, the energy in the effective channel matrix becomes more dispersed, with multiple non-zero elements emerging beyond the main peak and spreading across almost the entire matrix. Additionally, we observe that some rows or columns of the \( \mathbf{H}_i \) only contain very small values. This is due to the non-orthogonality introduced by bandwidth compression, which causes the ideal peak location $q^*$ to be non-integer, or even fall outside the allowable index range. In other words, for certain rows, no integer column index $q \in [0, N-1]$ enables $H_i[p, q]$ to reach its ideal peak value.

\section{IDFT-Based Signal Modulation}\label{Sec:SignalGen}
 This section first presents the implementation of the proposed nAFDM modulation using IDFT-based operations, followed by a complexity analysis showing that its computational cost is comparable to that of conventional orthogonal AFDM. 

\subsection{Signal Generation}
 In conventional AFDM systems, signal generation can be performed directly using a standard IDAFT. However, this is no longer feasible in nAFDM due to the loss of orthogonality, and direct implementation of \eqref{eq:Modu_Sym} incurs significant computational complexity. To address this challenge, we propose an efficient IDFT-based modulation scheme for nAFDM, where the input symbol vector is zero-padded and transformed via an IDFT module to generate the desired waveform.

 Specifically, we employ an IDFT of length $N' = \frac{N}{\alpha}$,  {where \(\alpha \in (0,1)\) is selected such that \(N'\in\mathbb{N}\) for practical FFT/IFFT implementation.}
The new input symbol vector ${\bf{x'}} \in \mathbb{C}^{N' \times 1}$ is defined as
\begin{align}
\label{eq:NewIput}
x'[m] = \begin{cases} 
x[m], & 0 \leq m< N, \\ 
0, & N \leq m < N',
\end{cases}
\end{align}
where the original data occupies the first $N$ positions, and the remaining $N' - N$ entries are zero-padded. The resultant waveform can be given by
\begin{align}
\label{eq:NonAFDM2}
    s'[n] = \frac{1}{\sqrt {N'}} \sum\limits_{m=0}^{N'-1} x'[m]\varphi_n(m), n = 0, \ldots, N'-1,
\end{align}
where $\varphi_n(m) = e^{\imath 2\pi\left ( c_1 n^2 + c_2 m^2 + \frac{nm}{N'} \right ) }$ denotes the chirp-based orthogonal basis function of {conventional AFDM.}
The transmitted signal is then obtained by truncating \(s'_{\alpha}[n]\) to its first \(N\) samples, discarding the remainder, and applying a normalization factor of $1/\sqrt{\alpha}$, which is expressed as 
\begin{align}
\label{eq:NonAFDM3}
    s_{\alpha}[n] = \frac{1}{\sqrt {\alpha}}s'[n], n = 0, \ldots, N-1.
\end{align}
 {{~~~\textit{Proof:} The detailed derivation of $s_{\alpha}$ is given in \textbf{Appendix B}.}} \hfill $\blacksquare$

 Notably, the proposed nAFDM framework can be configured to implement conventional AFDM, OFDM, OCDM, and SEFDM systems by appropriately selecting the parameters $c_1$, $c_2$, and $\alpha$. Specifically, setting $\alpha = 1$ and  $ c_1 = \frac{2(\alpha_{\max} + \xi_{\nu}) + 1}{2N}$ yields the orthogonal AFDM system, where the spacing factor \( \xi_{\nu} \) is used to mitigate fractional Doppler effects \cite{AFDM}; setting $\alpha = 1$ and $c_1 = c_2 = 0$ yields the orthogonal OFDM system; setting $\alpha = 1$ and $c_1 = c_2 = \frac{1}{2N}$ yields the orthogonal OCDM system;
and setting $\alpha\in(0,1)$ and $c_1 = c_2 = 0$ yields the non-orthogonal SEFDM system.

 \begin{figure}[htbp]
    \centering
    \begin{subfigure}{0.5\textwidth} 
        \centering
        \includegraphics[width=88mm]{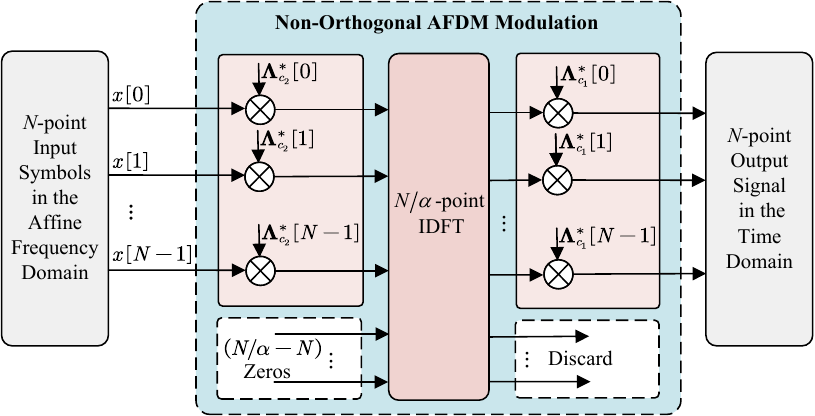}
        \subcaption{Modulation}
        \label{fig:Modulation}
    \end{subfigure}
    \hfill
    \vspace{1.5pt}
    \begin{subfigure}{0.5\textwidth}
        \centering
        \includegraphics[width=88mm]{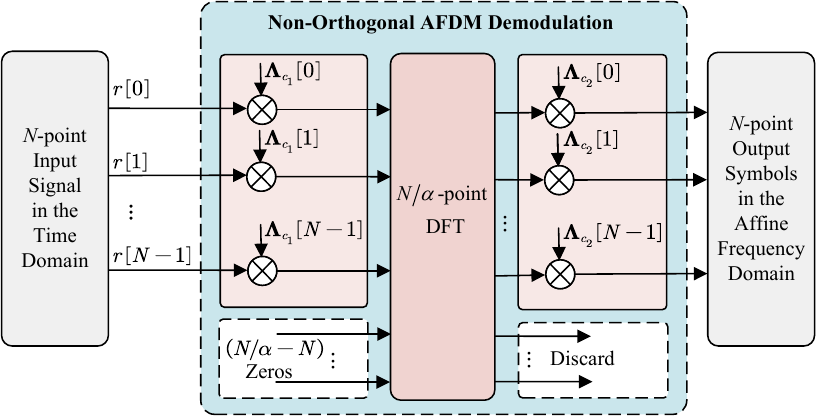}
        \subcaption{Demodulation}
        \label{fig:Demodulation}
    \end{subfigure}
    \caption{Modulation and demodulation processes of the proposed nAFDM scheme using IDFT and DFT operations: (a) Modulation; (b) Demodulation.}
    \label{fig:ModDemod_Block}
\end{figure}

Similarly, one can obtain the signal processing flow for the demodulation at the receiver. Fig.~\ref{fig:ModDemod_Block} illustrates the complete framework for nAFDM modulation and demodulation, including the signal transformations and mapping relationships throughout the transceiver.

\subsection{Complexity Analysis}
The main advantage of using IDFT-based signal generation is its significantly reduced computational complexity. By employing a simple zero-padding strategy and utilizing standard IDFT and DFT blocks of length $N'$, which are compatible with efficient radix-2 IFFT implementations, the complexity of nAFDM modulation is substantially reduced. Specifically, a direct implementation of the nAFDM scheme, as defined in \eqref{eq:Modu_Sym}, requires $N^2 + 2N$ complex multiplications and $N(N - 1)$ complex additions. In contrast, the proposed IDFT-based transmitter, implemented via an $N'$-point FFT with $N' = N/\alpha$, reduces the complexity to $\frac{N}{2\alpha} \log_2 \frac{N}{\alpha} + 2N$ complex multiplications and $\frac{N}{\alpha} \log_2 \frac{N}{\alpha}$ complex additions. Thus, the overall complexity of the nAFDM modulation is of the order $\mathcal{O}\left( \frac{N}{\alpha} \log_2\frac{N}{\alpha} \right)$. 
Furthermore, given the prior zero-padding operations, the IFFT computations {of} modulation contain several zero-input bins. This enables pruning of the IFFT trellis to eliminate redundant operations involving zero-operands \cite{SEFDM_SignalGen,NonAFDMGen}, thereby reducing the overall complexity to $\mathcal{O}\left( \frac{N}{\alpha} \log_2 N \right)$.



\renewcommand{\arraystretch}{1.6}
\begin{table}[htbp]    \caption{\MakeUppercase{Modulation complexity of OFDM, SEFDM, OCDM, AFDM, and proposed nAFDM systems}}
    \label{tab:gen_complexity} 
    \centering
    \footnotesize
    \begin{tabularx}{0.7\linewidth}{l | l} 
        \hline
        \textbf{Multi‑carrier systems} & \textbf{Order of complexity} \\ \hline
        OFDM & \(\mathcal{O}(N \log_2 N)\) \\ \hline
        SEFDM & \(\mathcal{O}\left(\tfrac{N}{\alpha} \log_2 N\right)\) \\ \hline
        OCDM & \(\mathcal{O}(N \log_2 N)\) \\ \hline
        AFDM & \(\mathcal{O}(N \log_2 N)\) \\ \hline
        Proposed nAFDM & \(\mathcal{O}\left(\tfrac{N}{\alpha} \log_2 {N}\right)\) \\ \hline
    \end{tabularx}
\end{table}
\renewcommand{\arraystretch}{1.0}

{Table~\ref{tab:gen_complexity} summarizes the modulation complexity
of OFDM, SEFDM, OCDM, AFDM, and the proposed nAFDM systems. OFDM rely on FFT-based implementations, whereas OCDM, AFDM and nAFDM involve additional steps beyond the FFT, which leads to a slight increase in computational complexity.} Compared to conventional orthogonal AFDM, which uses an $N$-point FFT,  nAFDM employs an extended transform length of $N' = N/\alpha$, leading to a moderate increase in FFT-related operations. Nevertheless, its overall complexity remains comparable to that of AFDM, while achieving improved SE. 
Moreover, the proposed design is compatible with conventional multi-carrier signaling frameworks, thus enhancing its practicality for bandwidth-limited and low-latency communication scenarios.

\section{ICI Analysis and Detection Design}\label{Sec:ICIAnalysis_Dec}
This section begins with {the ICI analysis of} the proposed nAFDM system, highlighting its dependence on the bandwidth compression factor. {Then, we provide} a brief overview of the MMSE detection algorithm and {discuss} its performance limitations under severe ICI. To improve detection performance, a soft ID algorithm is proposed that leverages symbol probabilities to progressively mitigate interference and refine symbol estimates. Subsequently, a low-complexity implementation strategy is introduced to reduce computational overhead while maintaining performance.

\subsection{ICI Analysis}\label{SubSec:ICIAnalysis}
Unlike conventional orthogonal systems, the proposed nAFDM waveform employs densely packed, partially overlapping chirp subcarriers to achieve higher SE. While this non-orthogonal structure enables bandwidth compression, it inevitably introduces ICI, which adversely affects signal detection. 
{In nAFDM,} the interference arising from non-orthogonally packed {chirp subcarriers} can be characterized by the modulation correlation matrix $\mathbf{C}_\alpha$, whose elements are defined as
\begin{align}
\label{eq:ICI1}
&C_ \alpha(m_1, m_2) = \sum_{n=0}^{N-1} A_\alpha(m_1, n) A_\alpha^*(m_2, n) \nonumber \\
&= \begin{cases}
1, & m_1 = m_2, \\
\frac{1}{N} e^{-\imath 2 \pi c_2 (m_1^2 - m_2^2)} F_\alpha(m_1,m_2), &  m_1 \neq m_2,
\end{cases}
\end{align}
where $F_\alpha(m_1, m_2)$ is given by
\begin{align}
\label{eq:ICI2}
F_\alpha(m_1,m_2)=\sum_{n=0}^{N-1}e^{-\imath\frac{2\pi\alpha}{N}(m_1-m_2)n}=\frac{1 - e^{-\imath 2\pi \alpha (m_1 - m_2)}}{1 - e^{-\imath \frac{2 \pi \alpha}{N} (m_1 - m_2)}}.
\end{align}
The matrix $\mathbf{C}_\alpha$ is Hermitian and reduces to a Toeplitz matrix when $c_2 = 0$. When $\alpha = 1$, {all the chirp subcarriers are orthogonal with each other,} and the correlation matrix reduces to an identity matrix. In contrast, for nAFDM systems with $\alpha < 1$, the off-diagonal elements of $\mathbf{C}_\alpha$ become nonzero, reflecting the presence of ICI among chirp subcarriers. The magnitude and distribution of these off-diagonal terms directly characterize the severity and structure of the resulting interference. Notably, when $\alpha(m_1 - m_2) \in \mathbb{Z}$, the corresponding correlation term $C_\alpha(m_1, m_2)$ becomes zero, implying orthogonality between certain pairs of chirp subcarriers even under non-orthogonal modulation.

\begin{figure}
    \centering
    \begin{subfigure}[b]{0.34\linewidth}
        \centering
        \includegraphics[width=\linewidth]{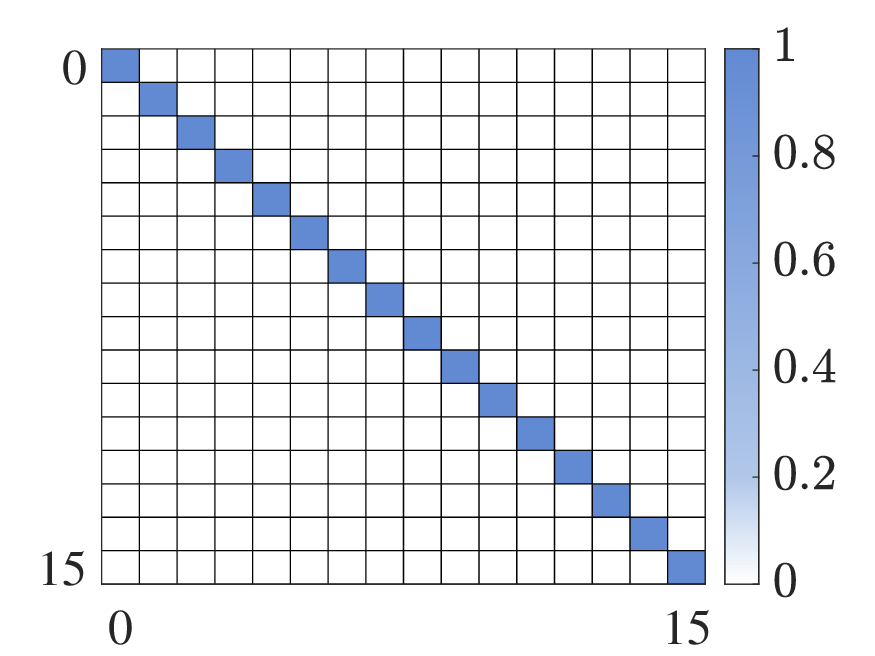}
        \caption{$\alpha=1$}
        \label{fig:ICI_Alpha100}
    \end{subfigure}
 \hspace*{-1em} 
    \begin{subfigure}[b]{0.34\linewidth}
        \centering
        \includegraphics[width=\linewidth]{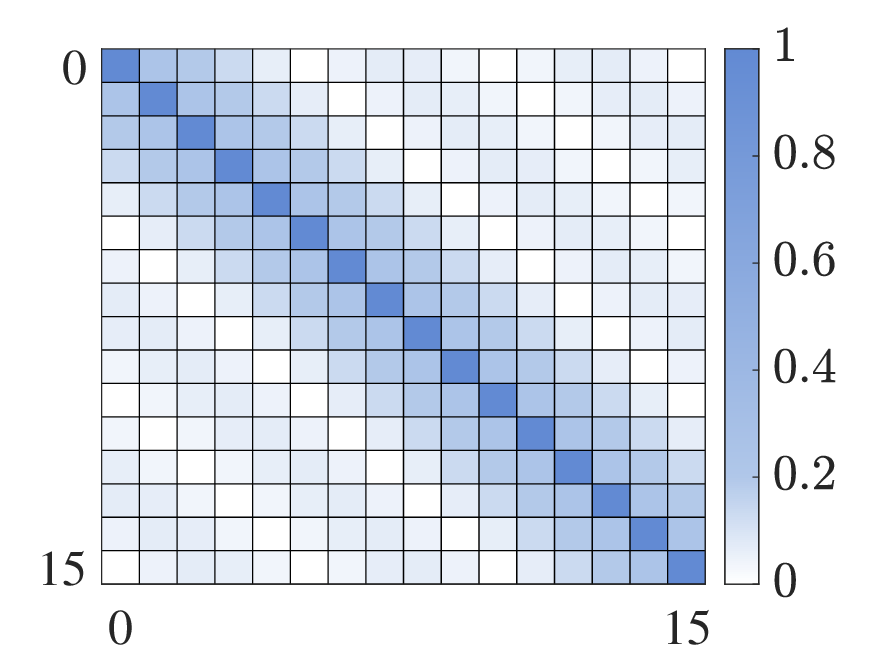}
        \caption{$\alpha=0.8$}
        \label{fig:ICI_Alpha80}
    \end{subfigure}
 \hspace*{-1em} 
    \begin{subfigure}[b]{0.34\linewidth}
        \centering
        \includegraphics[width=\linewidth]{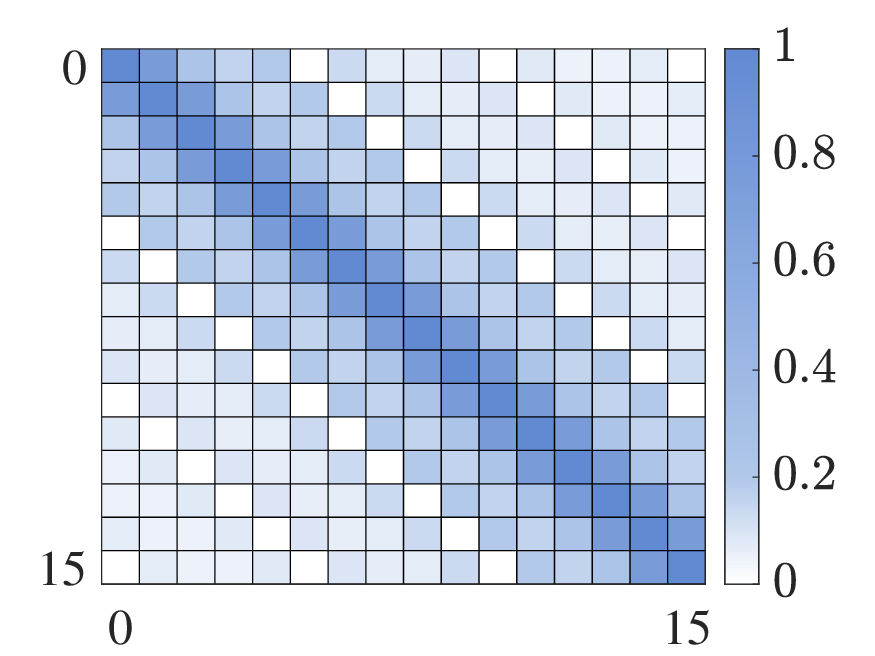}
        \caption{$\alpha=0.4$}
        \label{fig:ICI_Alpha40}
    \end{subfigure}
    \caption{{Structure of the correlation matrix $\mathbf{C}_\alpha$ for nAFDM under different compression factors ($N = 16$).}}
    \label{fig:ICIStructure}
\end{figure}

For example, when $N = 16$, ICI is eliminated at $|m_1 - m_2| = 5, 10, 15$ for $\alpha = 0.8$, {as observed} in Fig.~\ref{fig:ICIAmplitude} and Fig.~\ref{fig:ICIStructure}{(b)}, and at $|m_1 - m_2| = 10$ for $\alpha = 0.9$. This finding may offer new possibilities by allowing the proper selection of the bandwidth compression factor to increase the number of orthogonal subcarrier pairs, thereby enhancing bandwidth utilization, simplifying detection, and improving BER performance.

{To further analyze the impact of ICI, the magnitude of $C_ \alpha(m_1, m_2)$ can be expressed as 
\begin{align}
\label{eq:ICI3}
\left|C_ \alpha(m_1, m_2)\right|&=
\left|\frac{1}{N} e^{-\imath 2 \pi c_2 (m_1^2 - m_2^2)}F_\alpha(m_1,m_2) \right|\nonumber \\
&= \left|\frac{\sin(N\theta)}{N\sin(\theta)}\right|,
\end{align}
where $\theta \triangleq \frac{\pi}{N} \alpha(m_1 - m_2)$. 
\begin{figure}
    \centering
    \includegraphics[width=82mm]{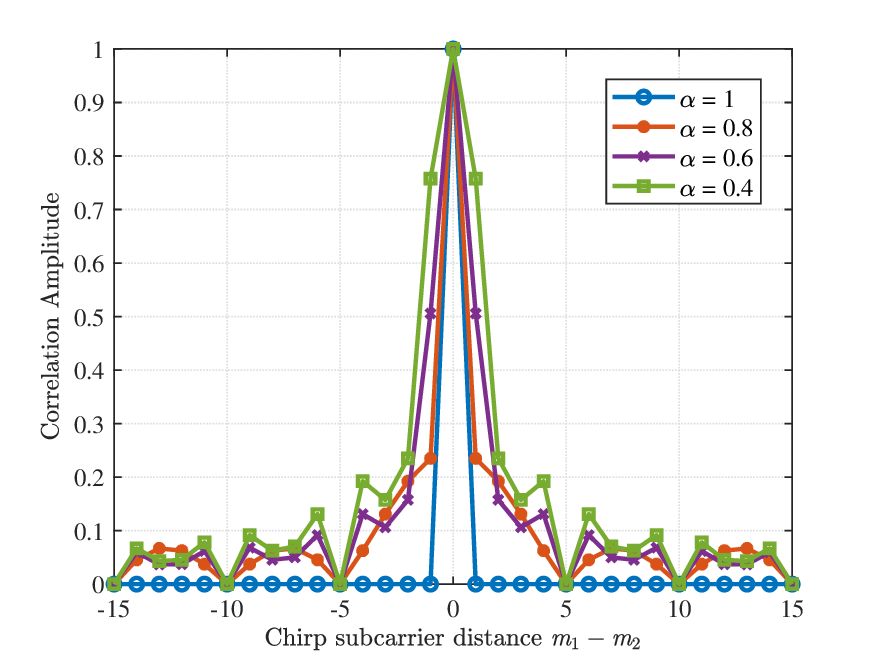}
    \caption{Correlation amplitude $\left|C_ \alpha(m_1, m_2)\right|$ for nAFDM under different compression factors ({$N = 16$)}.}
    \label{fig:ICIAmplitude}
\end{figure}
From \eqref{eq:ICI3}, it is clear that the cross-correlation depends on the bandwidth compression factor $\alpha$ and the relative separation between subcarriers $m_1$ and $m_2$. For a given $\alpha$, the correlation magnitude $|C_\alpha(m_1, m_2)|$ reaches its maximum at $m_1 = m_2$ and gradually decreases as the distance between subcarriers increases.}
{Fig.~\ref{fig:ICIAmplitude} illustrates the} correlation amplitude $\left|C_ \alpha(m_1, m_2)\right|$ in the proposed nAFDM {systems}. As the chirp subcarrier distance $|m_1 - m_2|$ increases, the correlation magnitude decreases, indicating that ICI is more severe between adjacent chirp subcarriers than distant ones. Moreover, the correlation amplitude is inversely proportional to $\alpha$, meaning that smaller $\alpha$ values lead to stronger interference among subcarriers and consequently degrade the system performance.



 Fig.~\ref{fig:ICIStructure} presents the structure of the correlation matrix $\mathbf{C}_\alpha$. When $\alpha = 1$, the matrix is strictly diagonal, reflecting perfect orthogonality. As $\alpha$ decreases, off-diagonal elements become more prominent, resulting in a denser and symmetric Hermitian matrix. Both Figs.~\ref{fig:ICIAmplitude} and~\ref{fig:ICIStructure} show how the compression factor affects both the magnitude and distribution of ICI in nAFDM, {thus offering valuable insights} for designing receiver algorithms that exploit this interference pattern for improved detection.




\subsection{MMSE Detection}
{We employ the MMSE detector in our proposed system, yielding the estimated time-domian signal as}
 \begin{align}\label{eq:mmse1}
\hat{\bf{s}}  =  (\mathbf{H}_\mathrm{T}^{{H}} \mathbf{H}_\mathrm{T} + \sigma^2 \mathbf{I}_N)^{-1}\mathbf{H}_{\mathrm{T}}^{{H}}\bf{r}.
\end{align}
By using the modulation matrix \( \mathbf{A}_\alpha \), the estimated signal is transformed from the time domain into the affine frequency domain {as}
\begin{align}
\label{eq:mmse4}
\bar{\mathbf{x}} = \mathbf{A}_\alpha \hat{\mathbf{s}} = \mathbf{C}_\alpha \mathbf{x} + \bar{\mathbf{w}},
\end{align}
where \( \bar{\mathbf{w}} \) represents the noise resulting from MMSE detection. {In AFDM systems}, since $\mathbf{C}_\alpha = \mathbf{I}$, the MMSE detector can directly recover the transmitted signal $\mathbf{x}$ from $\bar{\mathbf{x}}$, as the unitary modulation matrix does not introduce distortion. In contrast, for nAFDM systems, $\mathbf{C}_\alpha$ is non-diagonal, introducing dependencies between signal components and making the detection task more challenging. In this case, the linear MMSE detector may be insufficient to recover $\mathbf{x}$ accurately, as it is unable to effectively suppress interference introduced by the non-orthogonal modulation. Therefore, {we propose an advanced iterative interference cancellation strategy to enhance detection performance.}

\subsection{{Soft ID-Assisted} ICI Cancellation}\label{Sec:SoftId}

\begin{figure*}
    \centering
    \includegraphics[width=181mm]{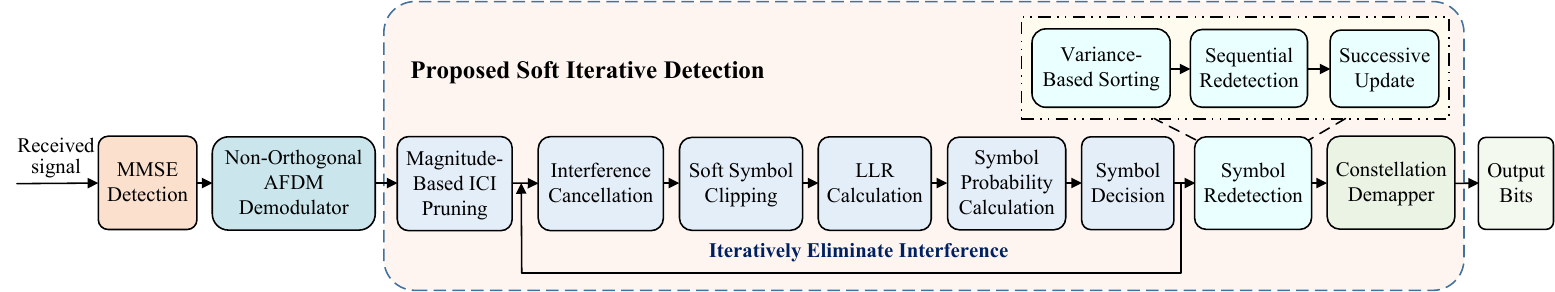}
    \captionsetup{subrefformat=parens} 
    \caption[left]{{Block diagram of the proposed soft ID algorithm for nAFDM systems.}}
    \label{fig:SoftIdProcess}
    \vspace{-1em}
\end{figure*}

From \eqref{eq:mmse4}, it can be seen that the signal \(\bf{\bar{x}}\) {is the distorted  transmitted signal} \(\bf{x}\) due to the ICI matrix \(\bf{C}_\alpha\).
Inspired by \cite{SEFDM_ID,NonAFDM_ID}, we propose a soft ID method to mitigate ICI and reconstruct the original transmitted signal, as illustrated in Fig.~\ref{fig:SoftIdProcess}. 
 {Specifically, the proposed method exploits symbol posterior probabilities to update the soft symbol estimates,  which are then used for progressive interference cancellation. Moreover, a selective redetection step is performed for certain selected low-reliability symbols according to their estimated variances and posterior probabilities, providing an additional refinement to the final decisions.}
The main steps of the proposed soft ID method are detailed as follows: 

\subsubsection{Initial Estimation Calculation}
The initial estimate ${\hat{\bar{\mathbf{x}}}}^{(0)}$ is obtained using {an} MMSE detector, i.e., \({\hat{\bar {\mathbf{x}}}}^{(0)}=\bar{\mathbf{x}}\). 

\subsubsection{Interference Cancellation}
At the $k$th iteration, the interference contributed by other subcarriers is estimated based on the detection result from the $(k-1)$th iteration. Specifically, the interference term is calculated as
    \begin{align} \label{eq:ICI}
        \mathbf{T}^{(k-1)} = (\mathbf{C}_\alpha - \mathbf{I}_N) 
        \hat{\bar{\mathbf{x}}}^{(k-1)},
    \end{align}
where \(\hat{\bar{\mathbf{x}}}^{(k-1)}
\) denotes the symbol vector recovered at iteration {\((k-1)\). To mitigate the estimated interference, we subtract \(\mathbf{T}^{(k-1)}\) from \(\bar{\mathbf{x}}\), yielding the refined signal}
\begin{align}\label{eq:RemoveICI}
\bar{\mathbf{z}}^{(k)} = \bar{\mathbf{x}} - \mathbf{T}^{(k-1)}.
\end{align}

\subsubsection{Soft Symbol Clipping}  
Since the soft symbols represent probability-weighted averages over constellation points, their values may deviate from the valid constellation region \cite{SoftIC}. Accordingly, a clipping operation is applied to constrain each element of \(\bar{\mathbf{z}}^{(k)}\) within the minimum and maximum bounds defined by the constellation. Specifically, for each component \(\bar z_n^{(k)}\), the real and imaginary parts are clipped separately as
\begin{align}\label{eq:SymbolClipp}
\Re\{z_n^{(k)}\} &= \min\left( \max\left( \Re\{\bar{z}_n^{(k)}\}, \mathcal{X}_\mathrm{min,real} \right), \mathcal{X}_\mathrm{max,real} \right),\nonumber \\
\Im\{z_n^{(k)}\} &= \min\left( \max\left( \Im\{\bar{z}_n^{(k)}\}, \mathcal{X}_\mathrm{min,imag} \right), \mathcal{X}_\mathrm{max,imag} \right),
\end{align}
 where \(\mathcal{X}_\mathrm{min,real}\), \(\mathcal{X}_\mathrm{max,real}\), \(\mathcal{X}_\mathrm{min,imag}\), and \(\mathcal{X}_\mathrm{max,imag}\) denote the minimum and maximum real and imaginary values among all {constellation points}, respectively. 

\captionsetup[figure]{singlelinecheck=off}
\captionsetup[figure]{singlelinecheck=on}

\subsubsection{{Log-likelihood Ratio (LLR) Calculation}}
After symbol clipping, the refined 
 symbol vector \(\mathbf{z}^{(k)}\) is utilized to compute the {LLRs} corresponding to the transmitted bits. Let \( M \) denote the modulation order, and each constellation point \( \mathcal{X}_m \in \mathcal{A} \) {is} mapped to a unique { \(\log_2M\)-bit} label according to Gray coding.  
For each symbol \( z_n^{(k)} \), the squared Euclidean distance to each constellation point is calculated as
\begin{align}\label{eq:LLR1}
d_n^{(k)}[m] = \left| z_n^{(k)} - \mathcal{X}_m \right|^2,  m = 1,2,\dots,M.
\end{align}
The LLR of the {$b$th bit} corresponding to symbol \( z_n^{(k)} \) is then given by
\begin{align}\label{eq:LLR}
\mathrm{LLR}_{b,n}^{(k)} = \log \left( \frac{\sum\limits_{\mathcal{X}_m \in \mathcal{S}_b^{0}} \exp\left( -\frac{d_n^{(k)}[m]}{2\sigma^2} \right)}{\sum\limits_{\mathcal{X}_m \in \mathcal{S}_b^{1}} \exp\left( -\frac{d_n^{(k)}[m]}{2\sigma^2} \right)} \right),
\end{align}
for $b = 1, \dots, \log_2M$, where \(\mathcal{S}_b^0\) and \(\mathcal{S}_b^1\) denote the sets of constellation points whose {\(b\)}th bit is 0 and 1, respectively.

\subsubsection{Symbol Probability Calculation}
The probabilities of each bit being 0 or 1 are calculated as
\begin{align}
P(b_n^{(k)}=0) &= \frac{1}{1+\exp\left(-\mathrm{LLR}_{b,n}^{(k)}\right)}, \\
P(b_n^{(k)}=1) &= 1 - P(b_n^{(k)}=0).
\end{align}
{Next, the probability of symbol \(z_n^{(k)}\) is mapped to a specific constellation point \(\mathcal{X}_m\), yielding}
\begin{align}\label{eq:Prob}
P_n^{(k)}(\mathcal{X}_m) = \prod_{b=1}^{\log_2(M)} P\left( b_n^{(k)} = l_b(\mathcal{X}_m) \right),
\end{align}
where \( l_b(\mathcal{X}_m) \in \{0,1\} \) denotes the {\(b\)}th bit of the label associated with \(\mathcal{X}_m\). The updated soft symbol \(\hat{z}_n^{(k)}\) is then computed as the probability-weighted average over all constellation points:
\begin{align}\label{SymbolSum}
\hat{z}_n^{(k)} = \sum_{m=1}^{M} \mathcal{X}_m \cdot P_n^{(k)}(\mathcal{X}_m).
\end{align}

\subsubsection{Symbol Decision}
{During the $k$th} iteration, the symbol $\hat{\bar{x}}[n]^{(k)}$ is updated by selecting the constellation point with the highest posterior probability, given by
\begin{align}\label{eq:UpdatedSym}
\hat{\bar{x}}[n]^{(k)} = 
\displaystyle\arg\max_{{\cal X}_m \in \mathcal{A}} P_n^{(k)}({\cal X}_m).
\end{align}
The resulting decision $\hat{\bar{x}}[n]^{(k)}$ is subsequently used for the interference cancellation process in the next iteration. To assess the reliability of the soft symbol estimate, the corresponding estimation variance is computed as
\begin{align}\label{eq:Var}
e_n^{(k)} = \sum_{m=1}^{M} \left| \mathcal{X}_m - \hat{z}_n^{(k)} \right|^2 \cdot P_n^{(k)}(\mathcal{X}_m),
\end{align}
{which} reflects the uncertainty associated with each soft estimate, {and} lower variance indicates higher reliability of the current symbol decision \cite{SoftIC}.

\subsubsection{Symbol Redetection}
After $K$ iterations, some soft symbol estimates may still exhibit low reliability, as indicated by their {estimated variances.} To further improve detection performance, a redetection procedure is introduced. This process jointly exploits both the estimated variances and the corresponding soft probabilities to selectively refine the final symbol decisions. The redetection process consists of the following steps:

\begin{itemize}
\item \textbf{{Variance-based} Sorting}: The estimation variances $\{e_0^{(K)}, e_1^{(K)}, \dotsc, e_{N-1}^{(K)}\}$ are sorted in descending order:
\begin{align}\label{GetUndeSymInd2}
e_{n_0}^{(K)} \geq e_{n_1}^{(K)} \geq \cdots \geq e_{n_{N-1}}^{(K)},
\end{align}
where $\{n_0, n_1, \dotsc, n_{N-1}\}$ denotes the sorted symbol indices. The top $|\mathcal{U}|$ symbols with the highest estimation variances are selected for redetection, where $\mathcal{U} = \{n_0, n_1, \dotsc, n_{|\mathcal{U}| - 1}\}$ denotes the set of selected indices.




\item \textbf{Sequential Redetection}:  
Starting from the symbol $\hat{\bar{x}}_{n_0}$ with the highest estimation variance, each symbol $\hat{\bar{x}}_{n_u}$ (for $n_u \in \mathcal{U}$) is sequentially redetected based on the mean squared error (MSE) criterion. For each $\hat{\bar{x}}_{n_u}$, we examine its $M{-}1$ alternative constellation candidates (excluding the current soft estimate), ranked in descending order of their associated posterior probabilities.
Among these candidates, the one minimizing the residual error is selected according to
\begin{align}\label{eq:MSE}
\hat{x}[n_u] = \arg \min_{{\cal X}_i \in \mathcal{A}} \left\| \mathbf{y} - \mathbf{H}_\mathrm{eff} \hat{\bar{\mathbf{x}}}_{[n_u \leftarrow {\cal X}_i]} \right\|^2,
\end{align}
where $\hat{x}[n_u]$ denotes the updated decision after redetection, and $\hat{\bar{\mathbf{x}}}_{[n_u \leftarrow {\cal X}_i]}$ is the tentative symbol vector with the {$n_u$th} entry replaced by candidate ${\cal X}_i$. {To reduce the complexity} of redetection, the residual in the MSE criterion can be efficiently updated by exploiting its incremental structure. Specifically, given the original residual \((\mathbf{y} - \mathbf{H}_\mathrm{eff} \hat{\bar{\mathbf{x}}})\), the updated residual in \eqref{eq:MSE} when replacing \(\hat{\bar x}_{n_u}\) with candidate \({\cal X}_i\) can be rewritten as
\begin{align}
\mathbf{y} - \mathbf{H}_\mathrm{eff} \hat{\bar{\mathbf{x}}}_{[n_u \leftarrow {\cal X}_i]} = (\mathbf{y} - \mathbf{H}_\mathrm{eff} \hat{\bar{\mathbf{x}}}) + \mathbf{h}_{n_u} ({\cal X}_i - \hat{\bar x}[n_u]).
\end{align}
where $\mathbf{h}_{n_u}$ represent the {$n_u$th} column of $\mathbf{H}_\mathrm{eff}$. Residual updates can thus be performed efficiently with linear complexity $N$ per candidate, resulting in an overall redetection complexity of approximately \(\mathcal{O}(N^2 + (M-1)N|\mathcal{U}|)\).

\item \textbf{Successive Update}: 
After each redetection step, the estimated symbol vector $\hat{\mathbf{x}}$ is updated by replacing the {$n_u$th} entry with the newly detected symbol, i.e.,
\begin{align}
{
\hat{\mathbf{x}}= \hat{\bar{\mathbf{x}}}_{[n_u \leftarrow \hat{x}[n_u]]},\quad u = 0,1,\dotsc,|\mathcal{U}|-1.}
\end{align}
The redetection procedure then proceeds to the next symbol $n_{u+1}$, and the update process is repeated sequentially until all symbols in $\mathcal{U}$ have been processed.
\end{itemize}

{The proposed} redetection strategy systematically reprocesses the {most unreliable} symbol decisions and refines them based on minimizing MSE criterion, thereby enhancing overall detection performance, especially in scenarios with strong interference and noise.

\subsection{Low-Complexity Implementation via ICI Pruning}\label{Sec:Low-com SoftId}

By fully exploiting the distributional properties of the ICI matrix analyzed in Subsection~\ref{SubSec:ICIAnalysis}, we propose a magnitude-based ICI pruning method to reduce the computational complexity of the soft ID algorithm by {only retaining} the dominant interference components.

\begin{figure}
    \centering
    \begin{subfigure}[b]{0.34\linewidth}
        \centering
        \includegraphics[width=\linewidth]{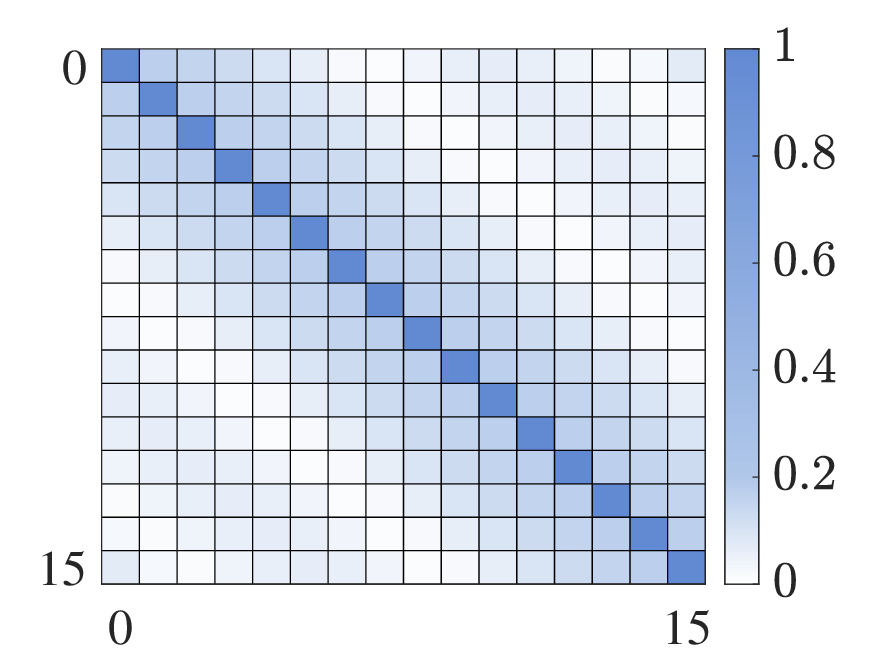}
        \caption{$D=N-1$}
        \label{fig:subplot1}
    \end{subfigure}
 \hspace*{-1em} 
    \begin{subfigure}[b]{0.34\linewidth}
        \centering
        \includegraphics[width=\linewidth]{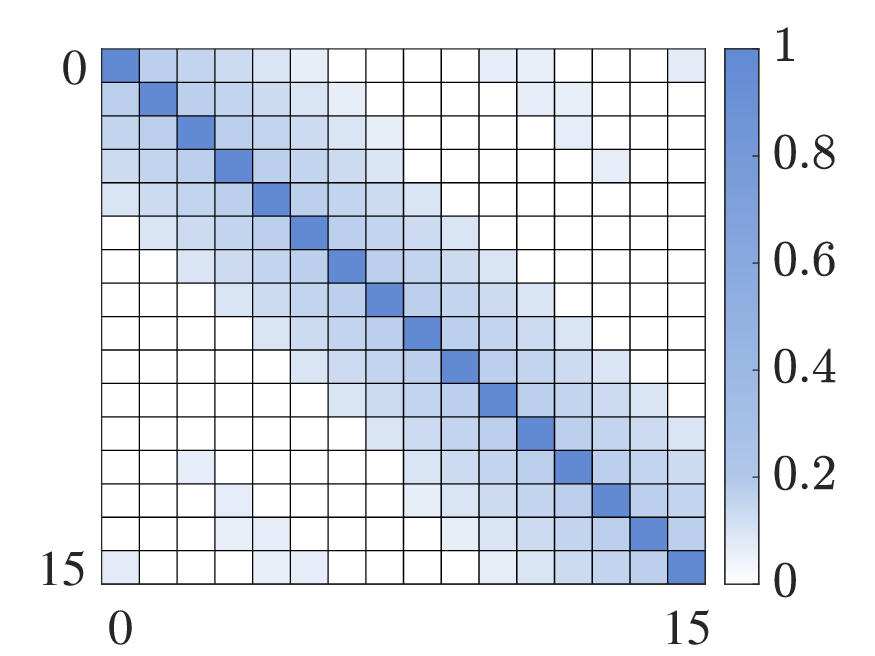}
        \caption{$D=\frac{N}{2}$}
        \label{fig:subplot2}
    \end{subfigure}
 \hspace*{-1em} 
    \begin{subfigure}[b]{0.34\linewidth}
        \centering
        \includegraphics[width=\linewidth]{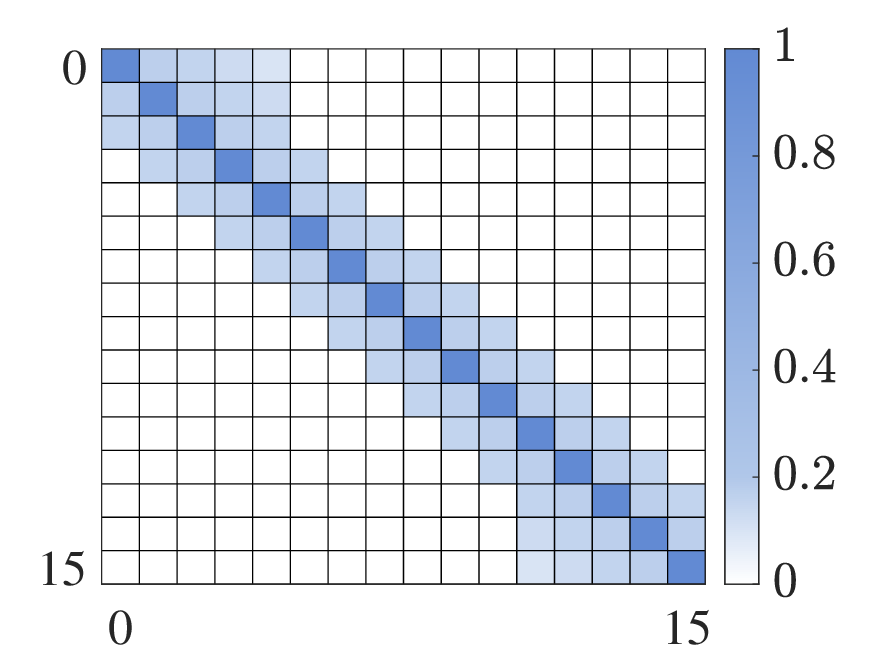}
        \caption{$D=\frac{N}{4}$}
        \label{fig:subplot3}
    \end{subfigure}
    \caption{{Truncated correlation matrix $\mathbf{C}_{\alpha,D}$ with $\alpha = 0.85$ and $N = 16$ under different ICI spans.}}
    \label{fig:ICIMatrix_DiffSpan}
\end{figure}

{We} first define $D$ as the ICI span, representing the number of dominant off-diagonal interference terms retained per chirp subcarrier. Let $\mathbf{C}_{\alpha,D}$ denote the truncated correlation matrix obtained by preserving, for each row $i$, the main diagonal elements and the $D$ largest off-diagonal elements (in magnitude). The construction is given by
\begin{align}\label{eq:CorrMatNew}
{C}_{\alpha,D}(i,j)
= 
\begin{cases}
{C}_{\alpha}(i,j), & j = i \text{ or } j \in \mathcal{S}_i, \\[3pt]
0, & \text{otherwise},
\end{cases}
\end{align}
where $\mathcal{S}_i$ denotes the set of indices corresponding to the $D$ largest off-diagonal entries in the {$i$th} row of $|\mathbf{C}_{\alpha}|$. {An illustration of the truncated correlation matrix $\mathbf{C}_{\alpha,D}$ with different values of $D$ is shown in Fig.~\ref{fig:ICIMatrix_DiffSpan}.
}

Under this approximation, the signal in \eqref{eq:mmse4} is rewritten as
\begin{align}\label{eq:LowSoftId1}
\bar{\mathbf{x}} = \mathbf{x}+(\mathbf{C}_{\alpha,D} -\mathbf{I}_N)\mathbf{x} + (\mathbf{C}_{\alpha} - \mathbf{C}_{\alpha,D})\mathbf{x} + \bar{\mathbf{n}},
\end{align}
where \( (\mathbf{C}_{\alpha,D} -\mathbf{I}_N)\mathbf{x}\) captures the dominant ICI components, while the residual interference term \( (\mathbf{C}_{\alpha} - \mathbf{C}_{\alpha,D})\mathbf{x} \) is treated as additional noise due to its relatively low energy. When \( D = 0 \), only the main diagonal elements of $\mathbf{C}_{\alpha}$ are retained, and all off-diagonal ICI terms are neglected,  yielding a purely diagonal approximation. In contrast, when \( D = N-1 \), all elements of \( \mathbf{C}_{\alpha} \) are preserved, corresponding to full ICI consideration. These represent the two extremes of the proposed magnitude-based ICI pruning approach.  
As \( D \) increases, more ICI components are explicitly mitigated, enhancing system performance at the cost of higher computational complexity. Therefore, $D$ enables a flexible trade-off between complexity and performance.

Accordingly, the subcarrier interference term in \eqref{eq:ICI} is reformulated as
\begin{align}\label{eq:ICINew} 
\mathbf{T}^{(k-1)} = (\mathbf{C}_{\alpha,D} - \mathbf{I}_N) \hat{\bar{\mathbf{x}}}^{(k-1)}.
\end{align}
The complete procedure of the proposed soft ID algorithm with magnitude-based ICI pruning is summarized in \textbf{Algorithm~\ref{alg:SoftId}}.


\begin{algorithm}
\caption{Proposed Soft ID Algorithm}
\label{alg:SoftId}
\begin{algorithmic}[1]
\footnotesize
\Require{Received signal $\mathbf{r}$, channel matrix $\mathbf{H}_\mathrm{T}$, modulation matrix $\mathbf{A}_{\alpha}$, number of iterations $K$, modulation order $M$, ICI span $D$.}
\Ensure{Detected symbol vector $\hat{\mathbf{x}}$.}

\State \textbf{Step 1: Initial Estimation}
\State Compute the initial estimate $\bar{\mathbf{x}}$ using \eqref{eq:mmse4}.

\State \textbf{Step 2: Magnitude-Based ICI Pruning}
\State Compute $\mathbf{C}_{\alpha,D}$ based on \eqref{eq:ICI1} and \eqref{eq:CorrMatNew}.

\State \textbf{Step 3: Iterative Interference Cancellation}
\For{$k = 1$ to $K$}
    \State Compute interference: $\mathbf{T}^{(k-1)} \leftarrow (\mathbf{C}_{\alpha,D} - \mathbf{I}_N) \hat{\bar{\mathbf{x}}}^{(k-1)}$.
    \State Interference cancellation: $\bar{\mathbf{z}}^{(k)} \leftarrow \bar{\mathbf{x}} - \mathbf{T}^{(k-1)}$.

    \For{$n = 1$ to $N$}
        \State Obtain $z_n^{(k)}$ according to \eqref{eq:SymbolClipp}.
        \State Compute $\mathrm{LLR}_{b,n}^{(k)}$ based on \eqref{eq:LLR}.
        \State Calculate $P_n^{(k)}(\mathcal{X}_m)$ via \eqref{eq:Prob}.
        \State Update $\hat{\bar{x}}_n^{(k)}$ using \eqref{eq:UpdatedSym}.
    \EndFor
\EndFor

\State \textbf{Step 4: Symbol Redetection}
\State Obtain $\mathcal{U}$ using \eqref{GetUndeSymInd2}.
\For{$u = 0$ to $|\mathcal{U}|-1$}
    \State Calculate $\hat{x}[n_u]$ based on \eqref{eq:MSE}.
\EndFor

\State \Return $\hat{\mathbf{x}}$.
\end{algorithmic}
\end{algorithm}

\subsection{Complexity Analysis}
The conventional MMSE detector exhibits a computational complexity of \(\mathcal{O}(N^2)\) \cite{MMSEComplex},  {while the ID scheme has a complexity order of \(\mathcal{O}(KN^{2})\) \cite{SEFDM_ID}.} The proposed soft ID algorithm, which builds upon the MMSE detection results, introduces additional iterative steps to mitigate subcarrier interference and enhance performance in nAFDM systems. As analyzed in Subsection~\ref{Sec:SoftId}, the number of complex multiplications required per iteration is $DN+6MN+MN\log_2 M$. Therefore, the total complexity of the proposed soft ID algorithm over \(K\) iterations, including the complexity of symbol redetection, can be expressed as {\(\mathcal{O}\left[N^2+\left(DN+MN+MN\log_2 M\right)K + MN|\mathcal{U}|\right]\)}.

\section{Simulation Results}\label{Sec:Results}
In this section, we present numerical results to evaluate the BER and SE performance of the proposed nAFDM systems. The maximum achievable SE is defined as \cite{SPE}
\begin{align}
\label{eq:SPE}
\eta_{\text{eff}} = \frac{r_c \cdot\log_2 M}{\alpha \left( 1 + \frac{T_{\text{CP}}}{T} \right)} \quad \text{(bits/s/Hz)},
\end{align}
where $r_c$ is the coding rate, with $r_c = 1$ representing an uncoded system and $r_c \in (0, 1)$ corresponding to a coded system, and \(T_{\text{CP}}\) denotes the CP duration. 

In all simulations, 4QAM modulation is adopted with \( N = 32 \) subcarriers, and the CP length is set to \( L_{\mathrm{CP}} = 8 \). The number of multipaths is set to \( P = 4 \), and the corresponding channel coefficients are modeled as \( h_i \sim \mathcal{CN}(0, 1/P) \). The maximum delay spread and normalized maximum Doppler shift are configured as \( l_{\max} = 3 \) and \( \nu_{\max} = 2 \), respectively. The Doppler shift for each path is generated using the Jake's model, given by \(\nu_i = \nu_{\max} \cos(\theta_i)\), where \(\theta_i \in [-\pi, \pi]\) follows a uniform distribution. 



\begin{figure}[htbp]
    \centering
    \begin{subfigure}{0.5\textwidth} 
        \centering
        \includegraphics[width=81mm]{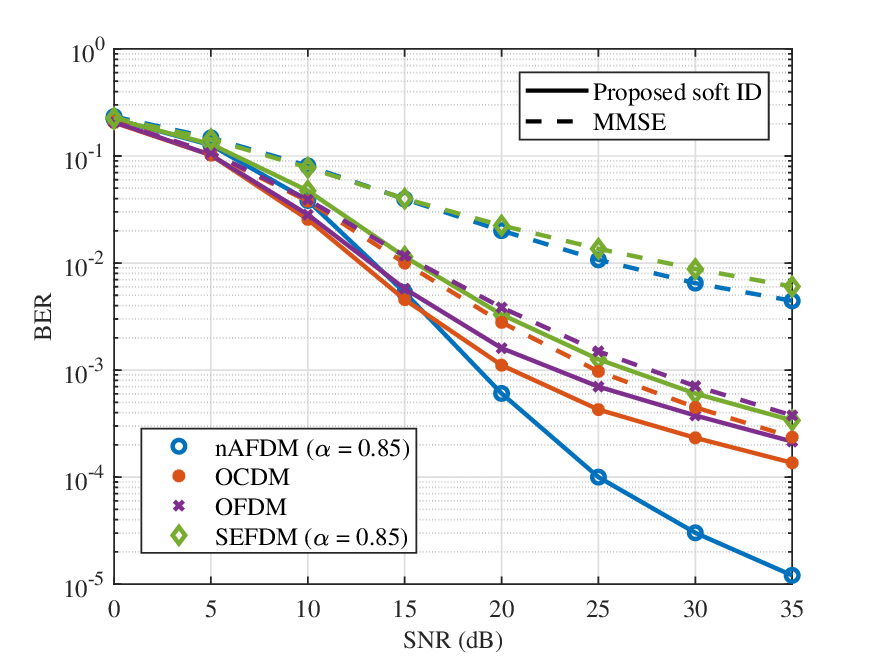}
        \subcaption{ {BER}}
        \label{fig:BER_DiffWaveform}
    \end{subfigure}
    \hfill
    \begin{subfigure}{0.5\textwidth}
        \centering
        \includegraphics[width=81mm]{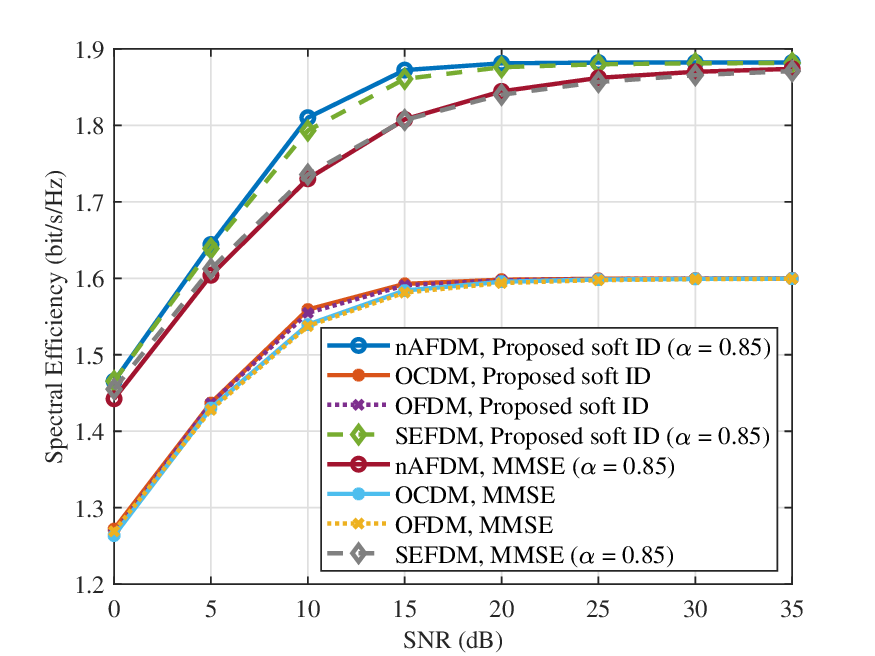}
        \subcaption{SE}
        \label{fig:SPE_DiffWaveform}
    \end{subfigure}
    \caption{BER and SE performance of nAFDM  {($\alpha=0.85$)}, OCDM, OFDM, and SEFDM using the MMSE and proposed soft ID schemes.}
    \label{fig:BER_SPE_DiffWaveform}
\end{figure}

\begin{figure}[htbp]
    \centering
    \begin{subfigure}{0.5\textwidth} 
        \centering
        \includegraphics[width=81mm]{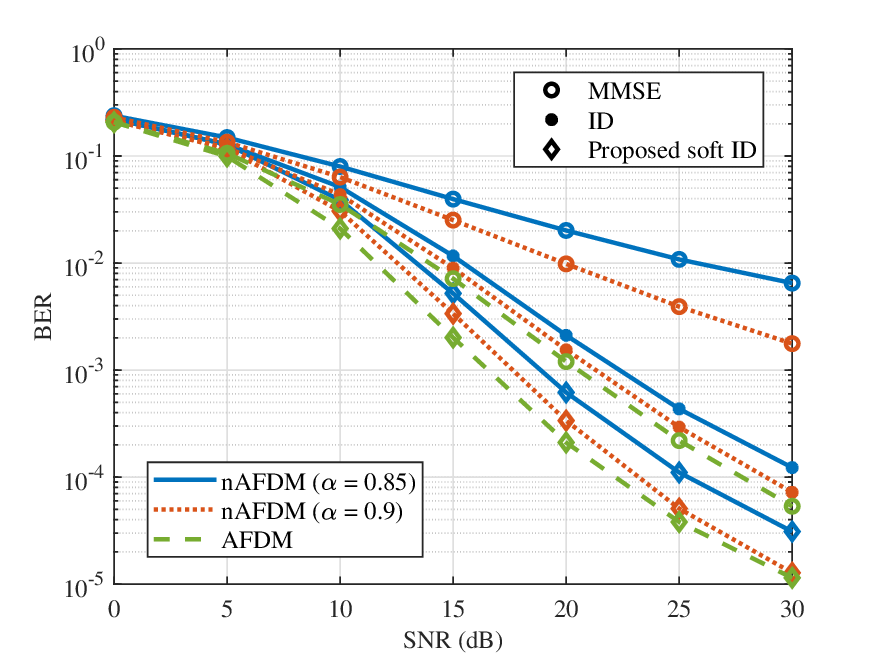}
        \subcaption{BER}
        \label{fig:BER_DiffDetection}
    \end{subfigure}
    \hfill
    \begin{subfigure}{0.5\textwidth}
        \centering
        \includegraphics[width=81mm]{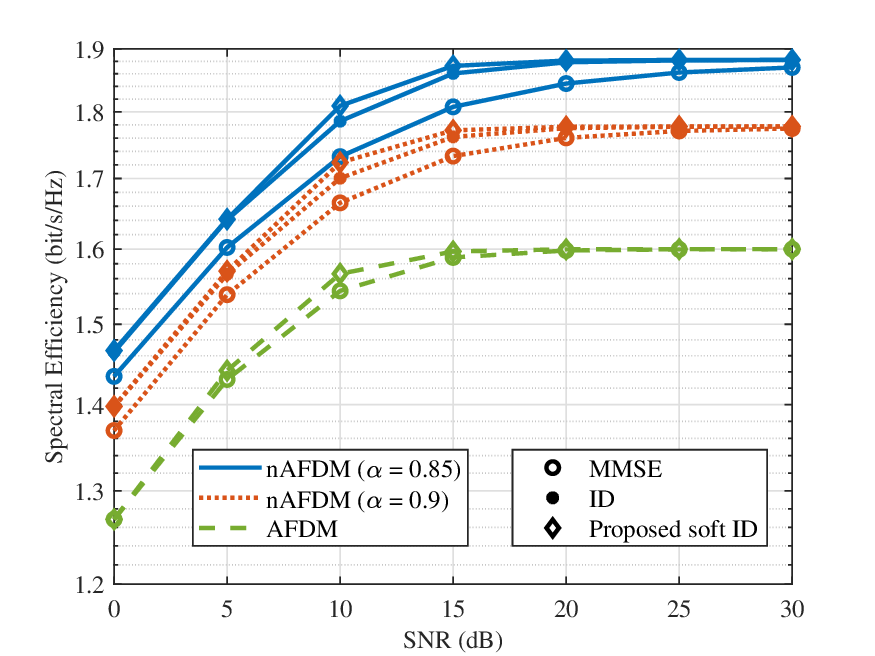}
        \subcaption{SE}
        \label{fig:SPE_DiffDetection}
    \end{subfigure}
    \caption{BER and SE performance of nAFDM and conventional AFDM systems using MMSE, ID, and the proposed soft ID schemes.}
    \label{fig:BER_SPE_DiffDetection}
\end{figure}

\subsection{Uncoded Waveform Comparison}

Fig.~\ref{fig:BER_SPE_DiffWaveform} presents the BER and {SE} of nAFDM, OCDM, OFDM, and SEFDM waveforms   using both MMSE and the proposed soft ID detection methods.
As shown in Fig.~\ref{fig:BER_SPE_DiffWaveform}{(a)}, nAFDM achieves significantly improved performance than SEFDM under the same bandwidth compression factor. This advantage arises from the chirp-based modulation nature of nAFDM, which provides a more uniform time-frequency distribution and stronger robustness to doubly selective channels. 
In contrast, SEFDM relies on conventional Fourier basis functions, making its BER more sensitive to Doppler. {However,} nAFDM exhibits worse performance than orthogonal OFDM and OCDM under MMSE detection, as the bandwidth compression destroys subcarrier orthogonality and introduces severe ICI that the linear MMSE equalizer struggles to work effectively. To address this limitation, we employ the proposed soft ID algorithm, which significantly enhances the performance of nAFDM and enables it to outperform OFDM and OCDM. Specifically, at a BER of $10^{-4}$, {nAFDM achieves a signal-to-noise ratio (SNR)} gain of over 10 dB {compared to OCDM and OFDM}. This improvement is mainly attributed to the iterative ICI cancellation in soft ID, which effectively mitigates the detrimental effects of non-orthogonality.
Moreover, at higher SNR levels, the availability of more reliable soft information further improves the effectiveness of soft ID, leading to substantial BER gains. {The effective SE calculated} based on BER values plotted in Fig.~\ref{fig:BER_SPE_DiffWaveform}{(a)}, is shown in Fig.~\ref{fig:BER_SPE_DiffWaveform}{(b)}. It can be seen that nAFDM achieves superior SE compared to both SEFDM and orthogonal waveforms such as OFDM and OCDM, {providing a 17.6$\%$ gain over OFDM and OCDM for SNR values higher than 15 dB.}

\subsection{Detection Method Comparison for Uncoded nAFDM}

Fig.~\ref{fig:BER_SPE_DiffDetection} compares the BER and {SE} of nAFDM and conventional AFDM using different detectors, including MMSE, ID \cite{SEFDM_ID}, and the proposed soft ID. Fig.~\ref{fig:BER_SPE_DiffDetection}{(a)} shows that both the ID and the proposed soft ID algorithms exhibit significant improvements over the conventional MMSE method, {thanks to the} iterative interference cancellation by exploiting the modulation correlation matrix. Moreover, the proposed soft ID algorithm outperforms the ID method by approximately 5 dB at a BER of $10^{-4}$, owing to its use of soft symbol information for enhanced iterative ICI cancellation. In addition, the {$\alpha = 0.9$ case} achieves superior BER performance compared to $\alpha = 0.85$, as a lower bandwidth compression factor $\alpha$ results in stronger ICI, leading to performance degradation. Furthermore, {given a BER of $10^{-4}$}, the proposed soft ID scheme for nAFDM yields approximately SNR gains of 2.4 dB ($\alpha = 0.85$) and 4.6 dB ($\alpha = 0.9$) over the MMSE algorithm for conventional AFDM, {respectively}. Notably, the proposed soft ID algorithm for nAFDM with $\alpha=0.9$ achieves {a nearly identical} BER performance {compared to} conventional AFDM.
In addition to BER performance, the proposed soft ID algorithm also {attains} notable improvements in effective SE, as shown in Fig.~\ref{fig:BER_SPE_DiffDetection}{(b)}. It achieves higher SE than both the ID and MMSE {counterparts}. Moreover, nAFDM consistently outperforms conventional AFDM in terms of SE.

\begin{figure}
    \centering
    \includegraphics[width=82mm]{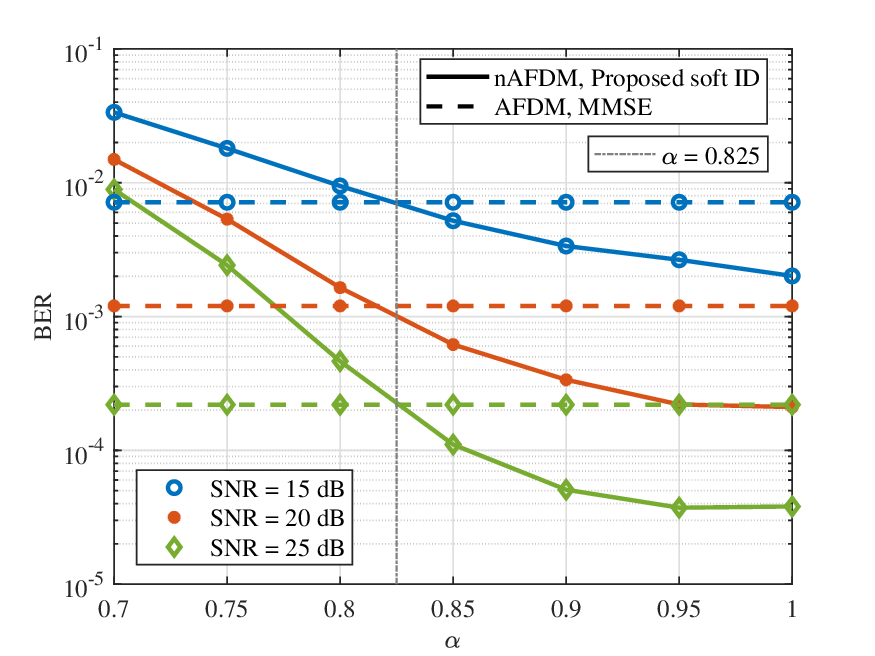}
    \caption{BER performance of the proposed soft ID scheme versus the bandwidth compression factor $\alpha$ at SNR = 15, 20, and 25 dB.}
    \label{fig:SoftIdDiffAlpha}
\end{figure}

{In Fig.~\ref{fig:SoftIdDiffAlpha},
we characterize the BER performance of nAFDM using the proposed soft ID and the conventional AFDM exploiting MMSE detector versus the bandwidth compression factor $\alpha$ under different SNR values.} As $\alpha$ increases, BER improves
due to less subcarrier compression. Notably, when $\alpha \ge 0.825$, the proposed soft ID scheme for nAFDM {is capable of achieving} BER performance comparable to or even better than {the MMSE detector} for AFDM.

\begin{figure}
    \centering
    \includegraphics[width=82mm]{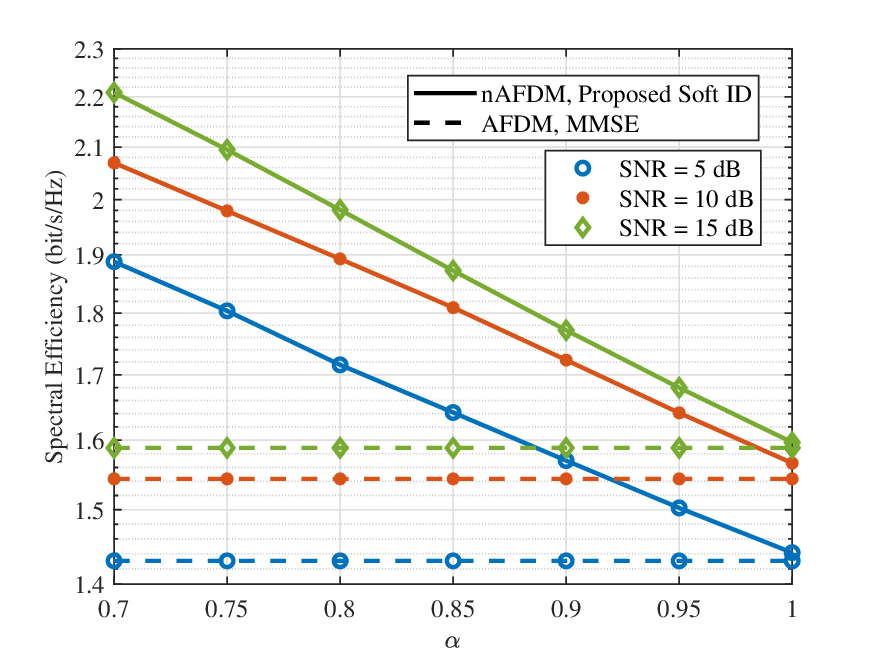}
    \caption{ {SE performance of the proposed soft ID scheme versus the bandwidth compression factor $\alpha$ at SNR = 5, 10, and 15 dB.}}
    \label{fig:SPE_SoftIdDiffAlpha}
\end{figure}

 {Fig.~\ref{fig:SPE_SoftIdDiffAlpha} further depicts the SE performance versus $\alpha$ for the proposed nAFDM with soft ID and the conventional AFDM with MMSE detector at SNR $=5$, $10$, and $15$ dB. It can be observed that the SE of the proposed nAFDM increases as $\alpha$ decreases, since a smaller $\alpha$ corresponds to stronger bandwidth compression and thus higher spectral utilization. In contrast, the SE of conventional AFDM remains constant with respect to $\alpha$ because it employs orthogonal subcarrier spacing (i.e., $\alpha = 1$). Moreover, the proposed nAFDM consistently achieves a higher SE than AFDM over the considered range of $\alpha$.}

{The above-mentioned} results indicate that the proposed soft ID scheme for nAFDM provides high flexibility, allowing $\alpha$ to be adjusted based on different communication requirements to achieve a better trade-off between BER and SE.



\begin{figure}
    \centering
    \includegraphics[width=82mm]{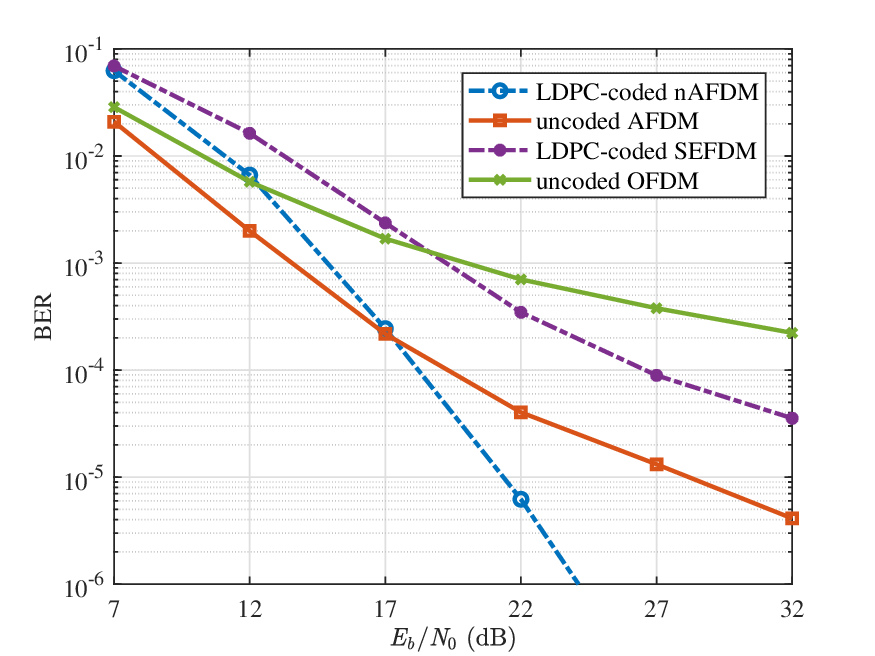}
    \caption{BER performance of a {rate-$5/6$} LDPC-coded nAFDM/SEFDM and uncoded AFDM/OFDM {using} the proposed soft ID scheme.}
    \label{fig:BER_codedNonAFDM_vs_AFDM}
\end{figure}

\subsection{Coded Waveform Comparison}
In this subsection, we present the BER and SE performance of the low-density parity-check (LDPC)-coded{\cite{LDPC}} nAFDM systems. {Fig.~\ref{fig:BER_codedNonAFDM_vs_AFDM}} illustrates the BER performance of coded nAFDM and SEFDM, as well as conventional uncoded AFDM and OFDM, all using the proposed soft ID scheme. For a fair comparison, both the code rate $r_c$ and the bandwidth compression factor $\alpha$ are set to $5/6$, yielding the same maximum achievable {SE} according to \eqref{eq:SPE}. The results show that when $E_b/N_0$ exceeds 18.6 dB, coded SEFDM outperforms uncoded OFDM. Furthermore, coded nAFDM surpasses conventional uncoded AFDM when $E_b/N_0$ is larger than 17 dB. Specifically, at a BER of $10^{-5}$, it achieves an $E_b/N_0$ gain of approximately 6.83 dB.
However, in low \( E_b/N_0 \) region, the BER performance of coded nAFDM is slightly worse due to the limited ICI cancellation capability of the soft ID under high noise conditions as well as the insufficient error correction capability at high code rates. In contrast, in high \( E_b/N_0 \) region, the noise impact is minimal, and the improved reliability of soft information enables more effective ICI mitigation for the proposed soft ID algorithm. Therefore, by combining nAFDM with channel coding, the system effectively mitigates ICI and enhances robustness against interference, resulting in superior BER performance.

\begin{figure}[htbp]
    \centering
    \begin{subfigure}{0.5\textwidth} 
        \centering
        \includegraphics[width=82mm]{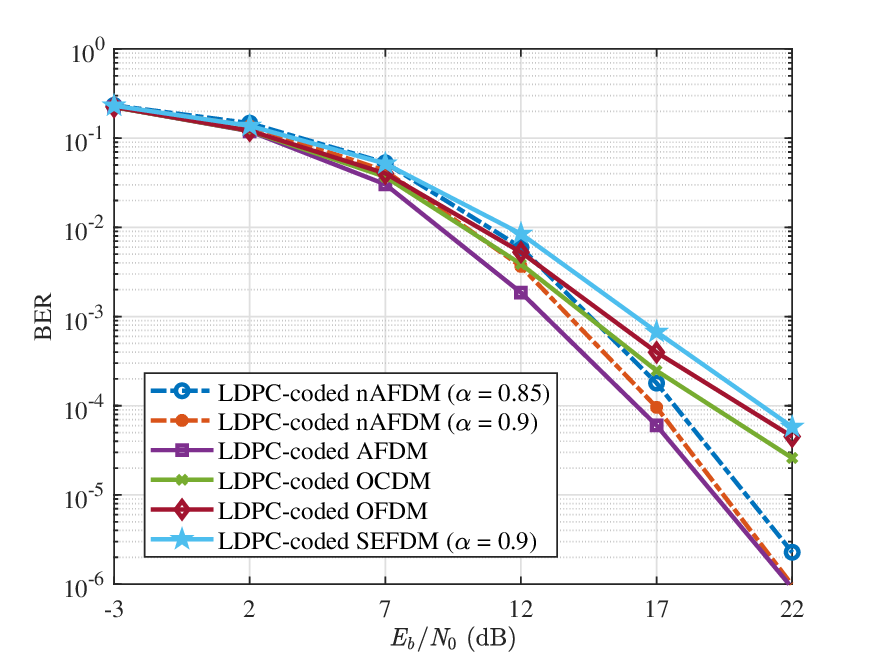}
        \subcaption{BER}
        \label{fig:BER_CodedDiffWaveform}
    \end{subfigure}
    \hfill
    \begin{subfigure}{0.5\textwidth}
        \centering
        \includegraphics[width=82mm]{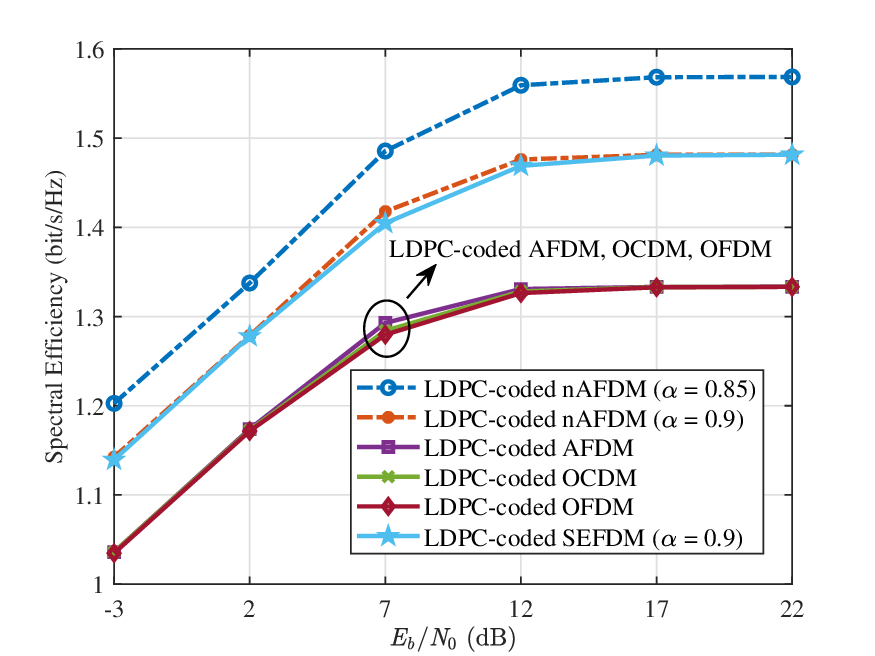}
        \subcaption{SE}
        \label{fig:SPE_CodedDiffWaveform}
    \end{subfigure}
    \caption{BER and SE performance of rate-$5/6$ LDPC-coded nAFDM, OCDM, OFDM, and SEFDM {using} the proposed soft ID scheme.}
    \label{fig:BER_SPE_CodedDiffWaveform}
\end{figure}

{Fig.~\ref{fig:BER_SPE_CodedDiffWaveform} shows the BLER and SE performance of rate-$5/6$ LDPC-coded nAFDM, OCDM, OFDM, and SEFDM waveforms using the proposed soft ID scheme. As depicted in Fig.~\ref{fig:BER_SPE_CodedDiffWaveform}(a), nAFDM with $\alpha = 0.9$ achieves BER performance comparable to that of AFDM, while significantly outperforming conventional orthogonal waveforms such as OCDM and OFDM, as well as the non-orthogonal SEFDM. Specifically, at a BER of $10^{-4}$, nAFDM with $\alpha=0.9$ yields performance gains of approximately 2.06 dB, 3.22 dB, and 3.94 dB over OCDM, OFDM, and SEFDM, respectively.
According to the BER results in Fig.~\ref{fig:BER_SPE_CodedDiffWaveform}(a), Fig.~\ref{fig:BER_SPE_CodedDiffWaveform}(b) further illustrates the effective SE. It can be seen that nAFDM offers a substantial SE advantage over conventional AFDM, OCDM, and OFDM. 
Moreover, these observations are consistent with the uncoded performance presented in Figs.~\ref{fig:BER_SPE_DiffWaveform} and ~\ref{fig:BER_SPE_DiffDetection}, confirming the superiority of nAFDM across both coded and uncoded scenarios.}


\begin{figure}
    \centering
    \begin{subfigure}{0.5\textwidth} 
        \centering
        \includegraphics[width=81mm]{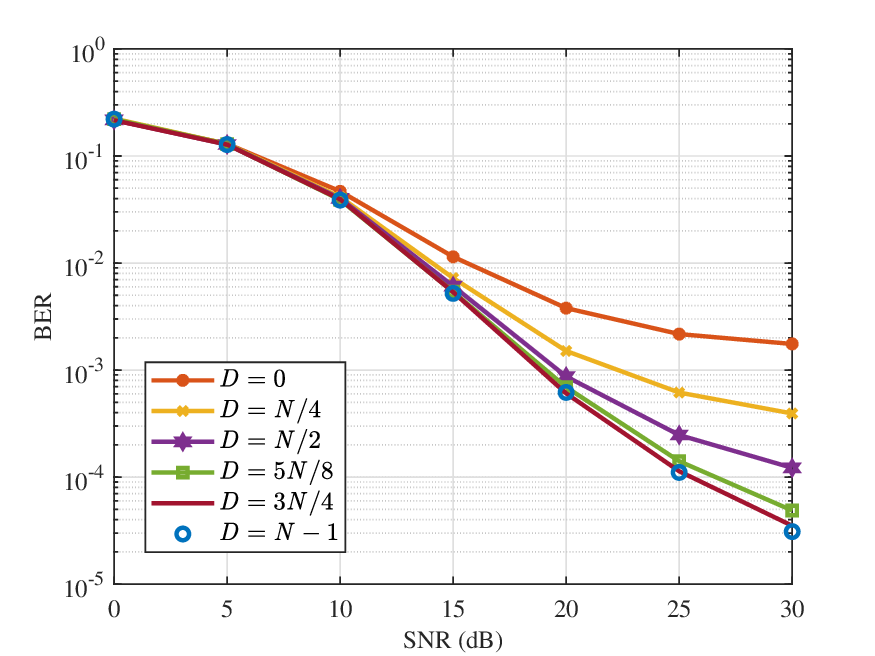}
        \subcaption{$\alpha=0.85$}
        \label{fig:85SoftIdICISpan}
    \end{subfigure}
    \hfill
    \begin{subfigure}{0.5\textwidth}
        \centering
        \includegraphics[width=81mm]{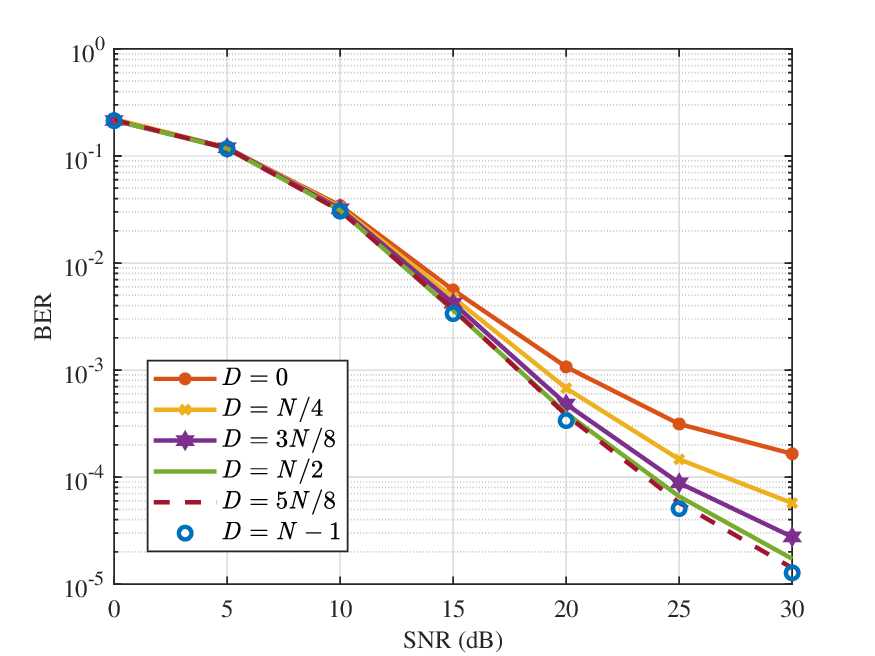}
        \subcaption{$\alpha=0.9$}
        \label{fig:90SoftIdICISpan}
    \end{subfigure}
    \caption{{BER performance of the proposed soft ID scheme for nAFDM with $\alpha=0.85$ and $\alpha=0.9$ under different ICI span \(D\).}}
    \label{fig:85_90_SoftIdICISpan}
\end{figure}

\subsection{Low-Complexity Soft ID Performance for Uncoded nAFDM}
All previous simulation results of {the proposed} soft ID scheme are based on the full ICI consideration without pruning {i.e., \( D = N-1 \)}. This subsection evaluates the BER performance of the proposed soft ID algorithm under varying ICI spans \(D\), demonstrating the effect of magnitude-based ICI pruning strategy on detection performance. 

Fig.~\ref{fig:85_90_SoftIdICISpan} shows the BER performance of {the proposed} soft ID for nAFDM with $\alpha=0.85$ and $\alpha=0.9$ under different values of \(D\). The case \(D = N - 1\) represents full ICI cancellation and serves as the {benchmark}. It can be observed that when \(D = 0\), the soft ID exhibits poor performance due to the complete neglect of ICI,  leaving substantial residual interference unmitigated. As \(D\) increases, more ICI components are incorporated into the iterative interference cancellation process, progressively improving the BER performance. Specifically, for \(\alpha = 0.85\) with \(D = \tfrac{3N}{4}\), the BER is nearly {identical compared to the} full ICI cancellation {scenario}, while reducing the ICI cancellation complexity by approximately 25\%. Similarly, for \(\alpha = 0.9\) with \(D = \tfrac{5N}{8}\), the BER approaches that of the full ICI case, with an ICI cancellation complexity reduction of approximately 37.5\%. Since the bandwidth compression factor $\alpha$ directly influences the ICI level, a smaller $\alpha$ induces stronger interference, requiring a larger ICI span $D$ to achieve satisfactory performance. For a given $\alpha$, increasing $D$ enhances ICI suppression and improves BER, but at the cost of higher computational complexity. Notably, a moderately reduced $D$ can significantly lower complexity while only minimal BER loss. These results demonstrate that choosing an appropriate ICI span $D$ enables a flexible {complexity vs. BER} trade-off, providing practical guidelines for low-complexity nAFDM system implementation.


\subsection{ {Uncoded nAFDM Performance Under Imperfect Channel State Information (CSI) and Extended Vehicular A (EVA) Channels}}

\begin{figure}
    \centering
    \includegraphics[width=82mm]{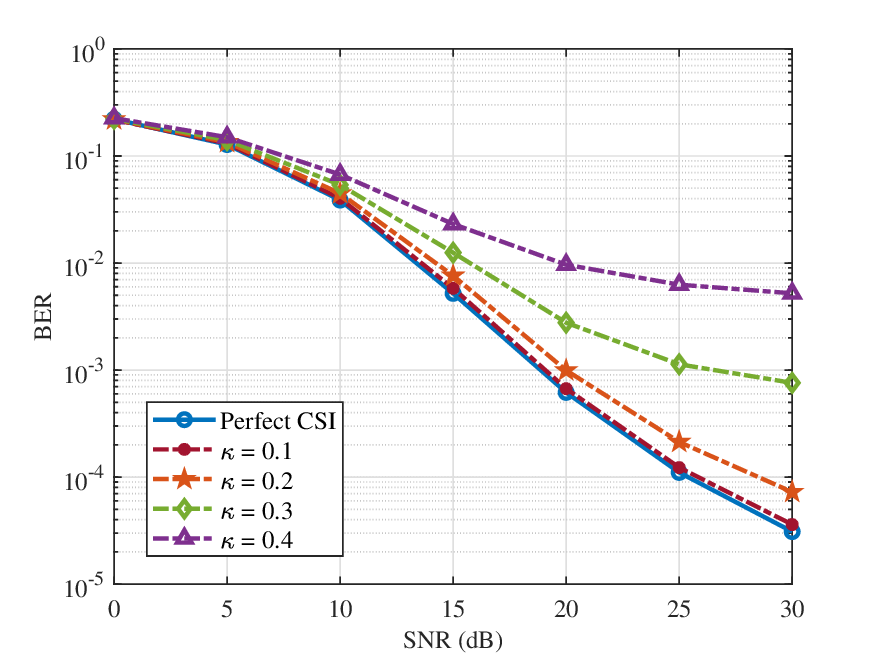}
    \caption{ {BER performance of nAFDM with $\alpha = 0.85$ using the proposed soft ID scheme with different values of the normalized channel estimation error coefficient $\kappa$.}}
    \label{fig:BER_ImperfectCSI}
\end{figure}

 {In this subsection, we evaluate the BER performance of the proposed scheme in the presence of imperfect CSI. Specifically, the imperfect channel estimation is modeled as $\hat{\mathbf{h}} = \mathbf{h}\left(1+\kappa \sigma_h\right)$, where $\mathbf{h}=[h_1,h_2,\ldots,h_P]$ is the channel coefficient vector, $\kappa \in (0,1)$ denotes the normalized channel estimation error coefficient, and the complex-valued random variable $\sigma_h$ follows a uniform distribution over the unitary circle \cite{ImperfectCSI_R1Add}. Fig.~\ref{fig:BER_ImperfectCSI} illustrates the BER performance of the nAFDM with $\alpha=0.85$ employing the proposed soft ID scheme for different values of $\kappa$, where $\kappa$ is set to 0.1, 0.2, 0.3, and 0.4, respectively. As expected, a larger $\kappa$ leads to a more pronounced BER degradation due to the increased channel mismatch. 
More specifically, the BER curve for $\kappa=0.1$ is nearly identical to that under perfect CSI.
At a target BER of $10^{-3}$, $\kappa=0.2$ incurs only a marginal SNR loss of about 1.1~dB, while $\kappa=0.3$ results in a more noticeable SNR loss of about 7.7~dB. Moreover, when $\kappa=0.4$, an evident error floor appears at around $5\times10^{-3}$, as the residual channel estimation error dominates the detection performance in the high SNR regime.}

\begin{figure}
    \centering
    \begin{minipage}{0.245\textwidth} 
        \centering
        \includegraphics[width=\linewidth]{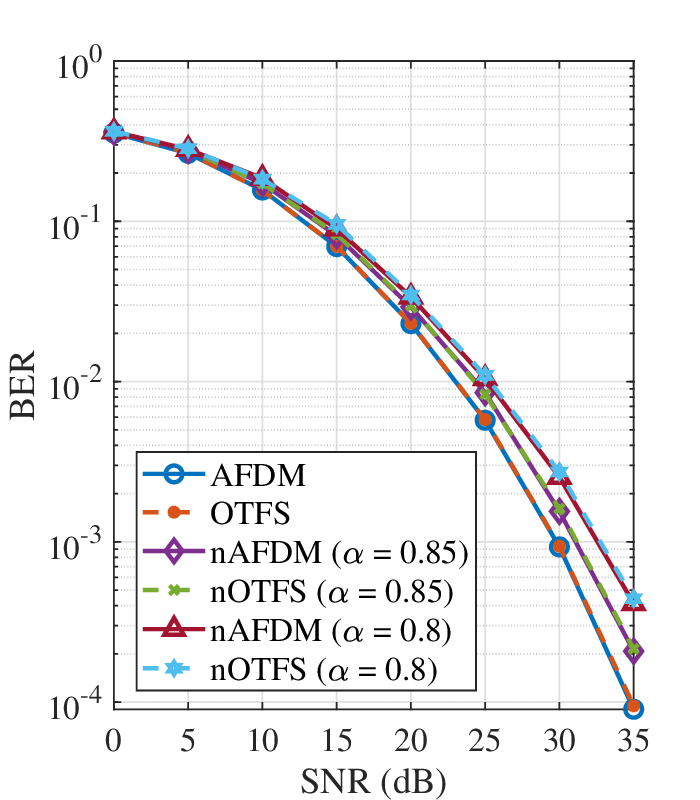}
        \subcaption{BER}
        \label{fig:BER_NonAFDM_NonOTFS}
    \end{minipage}
    \hspace*{-1.2em} 
    \begin{minipage}{0.245\textwidth}
        \centering
        \includegraphics[width=\linewidth]{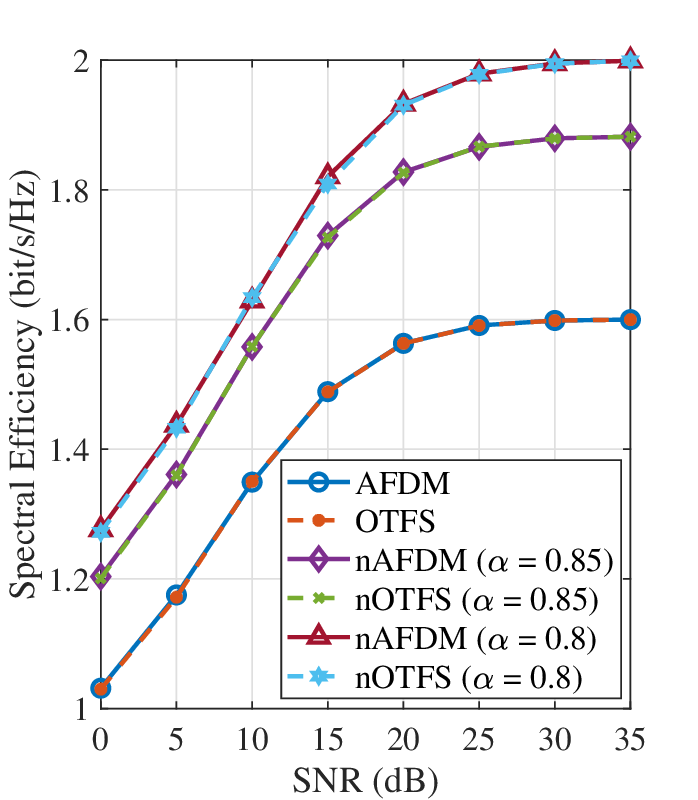}
        \subcaption{SE}
        \label{fig:fig:SPE_NonAFDM_NonOTFS}
    \end{minipage}
    \captionsetup{subrefformat=parens} %
    \caption{ {BER and SE performance of AFDM, nAFDM, OTFS and nOTFS using the proposed soft ID scheme under the EVA channel model.}}
    \label{fig:BER_SPE_NonAFDM_NonOTFS}
\end{figure}

 {To further evaluate the proposed nAFDM against its OTFS-based counterpart, we consider a spectrally efficient non-orthogonal OTFS (nOTFS) variant constructed by introducing the bandwidth compression factor $\alpha$ into the conventional OTFS modulation framework. This bandwidth compression induces non-orthogonality along the frequency (delay) dimension while preserving orthogonality along the time (Doppler) dimension.
Fig.~\ref{fig:BER_SPE_NonAFDM_NonOTFS} compares the BER and SE of AFDM, nAFDM, OTFS, and nOTFS using the proposed soft ID scheme under the EVA channel model \cite{EVA_Channel_R1Add}.
For a fair comparison,
 we set $M_{\mathrm d}=4$ delay bins and $N_{\mathrm d}=8$ Doppler bins for OTFS and nOTFS,
 satisfying \(M_{\mathrm d}N_{\mathrm d}=N\). Moreover, both nAFDM and nOTFS use the same bandwidth compression factors \(\alpha\in\{0.85,0.8\}\).
Fig.~\ref{fig:BER_SPE_NonAFDM_NonOTFS}(a) shows that AFDM and OTFS achieve the best BER performance, while nAFDM and nOTFS exhibit a slight BER degradation due to the additional ICI introduced by bandwidth compression. Moreover, reducing the compression factor from $\alpha=0.85$ to $\alpha=0.8$ leads to stronger ICI and hence worsens the BER performance for both nAFDM and nOTFS systems. The effective SE is further illustrated in Fig.~\ref{fig:BER_SPE_NonAFDM_NonOTFS}(b). It demonstrates that both nAFDM and nOTFS significantly improve the effective SE compared with their orthogonal counterparts, and a smaller $\alpha$ yields a higher SE. Moreover, the above results show that both nAFDM and nOTFS are capable of achieving nearly identical BER and SE, while nAFDM exhibits a lower modulation complexity\footnote{ {nOTFS modulation comprises a two-dimensional ISFFT operation followed by a frequency-compressed IFFT operation, yielding a complexity of $\mathcal{O}\left(N\log_2 N+\frac{N}{\alpha}\log_2 M_{\mathrm d}\right)$. In contrast, the proposed nAFDM modulation exhibits a complexity of $\mathcal{O}\left(\frac{N}{\alpha}\log_2 N\right)$, as summarized in Table~\ref{tab:gen_complexity}.   
To provide a quantitative complexity comparison, we consider two representative parameter settings. For $(N,M_{\mathrm d},N_{\mathrm d},\alpha)=(32,4,8,0.8)$, the nOTFS and proposed nAFDM modulators require about 240 and 200 complex multiplications, respectively. Furthermore, for a larger frame size with $(N,M_{\mathrm d},N_{\mathrm d},\alpha)=(1024,32,32,0.8)$, the multiplication counts increase to approximately 16640 for nOTFS and 12800 for nAFDM, respectively.}}.
}

\section{Conclusion}\label{Sec:Conclusion}
In this paper, we have proposed a novel nAFDM waveform for high-mobility communications with higher SE. {Compared to conventional AFDM,} the proposed waveform retains the advantageous affine structure of traditional AFDM while relaxing orthogonality to improve spectral utilization. We have first established the signal model and derived the input-output {signal} relationship, emphasizing the unique characteristics of the proposed waveform. Next, an efficient signal generation scheme based on IDFT has been introduced, demonstrating that the proposed nAFDM system can be practically implemented using existing IFFT and FFT processing modules without additional hardware complexity. {Moreover, we have} analyzed the ICI introduced by non-orthogonal modulation and derived its closed-form dependence on the bandwidth compression factor. To mitigate this interference, a soft ID algorithm has been developed, along with a low-complexity implementation strategy that leverages the distribution characteristics of ICI. Simulation results demonstrate that the proposed nAFDM waveform outperforms non-orthogonal SEFDM and conventional orthogonal schemes such as OFDM and OCDM in both BER and SE. More importantly, it achieves BER performance close to that of traditional orthogonal AFDM while significantly improving {SE}.
These results indicate the feasibility and advantages of the proposed nAFDM framework, making it a promising candidate for next-generation wireless systems, especially in high-mobility environments demanding spectral agility.
\section*{Appendix A \\ Derivation of the Subchannel Matrix \(\mathbf{H}_i\)}
The subchannel matrix \(\mathbf{H}_i\) in \eqref{eq:RxSignal_AFD2} is equivalently written as
\begin{align}\label{eq:H_i}
\mathbf{H}_i = \mathbf{\Lambda}_{c_2}
\underbrace{\mathbf{F}_\alpha
\overbrace{\mathbf{\Lambda}_{c_1}\mathbf{\Delta}_{f_i}\boldsymbol{\Pi}^{l_i}\mathbf{\Lambda}_{c_1}^H}^{\mathbf{V}}
\mathbf{F}_\alpha^H}_{\mathbf{W}}
\mathbf{\Lambda}_{c_2}^H,
\end{align}
where \(\mathbf{W} = \mathbf{F}_\alpha \mathbf{V} \mathbf{F}_\alpha^H\) and \(\mathbf{V} = \mathbf{\Lambda}_{c_1}\mathbf{\Delta}_{f_i}\boldsymbol{\Pi}^{l_i}\mathbf{\Lambda}_{c_1}^H\).
The \((n,m)\) element of \(\mathbf{V}\) can be written as
\begin{align}\label{eq:V1_new}
{V}[n,m] = {\Lambda}_{c_1}[n,n]\, { \Delta}_{f_i}[n,n]\, {\Pi}^{l_i}[n,m]\, \mathbf{\Lambda}_{c_1}^*[m,m].
\end{align}
According to the structure of \(\boldsymbol{\Pi}\) in \eqref{eq:Chan_ForCycShiftMat}, we have
\begin{align}\label{eq:Pi_entry}
{\Pi}^{l_i}[n,m] =
\begin{cases}
1,& m = (n-l_i)_N,\\
0,& \text{otherwise}.
\end{cases}
\end{align}
Substituting \eqref{eq:Pi_entry} into \eqref{eq:V1_new} yields
\begin{align}\label{eq:V_final_new}
{V}[n,m] =
\begin{cases}
e^{\imath 2\pi c_1 l_i^2} e^{-\imath 2\pi n(f_i + 2c_1 l_i)}, & m = (n-l_i)_N,\\
0, & \text{otherwise}.
\end{cases}
\end{align}
Next, the \((p,q)\) entry of \(\mathbf{W}\) is given by
\begin{align}\label{eq:W_expand}
{W}[p,q] = \sum_{n=0}^{N-1}\sum_{m=0}^{N-1} {F}_\alpha[p,n]\, {V}[n,m]\, {F}_\alpha^*[m,q].
\end{align}
Putting \eqref{eq:V_final_new} into \eqref{eq:W_expand} leads to
\begin{align}\label{eq:W_result}
{W}[p,q] = \frac{1}{N} e^{{\imath2\pi}c_1l_i^2 } U[p,q],
\end{align}
where
\begin{align}\label{eq:I_sum_new}
U[p,q] = \sum_{n=0}^{N-1} e^{-\frac{\imath2\pi}{N}\left[\alpha p n + N\left(f_i+2c_1l_i\right)n - \alpha q \left(\left(n-l_i\right)_N\right)\right]}.
\end{align}
Here, the modulo operation \((n-l_i)_N\) can be written as
\begin{align}\label{eq:modulo_shift}
(n-l_i)_N =
\begin{cases}
n-l_i+N, & 0 \leq n < l_i,\\
n-l_i, & l_i \leq n < N.
\end{cases}
\end{align}
Applying \eqref{eq:modulo_shift} simplifies \(U[p,q]\) to
\begin{align}\label{eq:I_sum_simplified}
U[p,q] =e^{-\imath\frac{2\pi\alpha q l_i}{N}} \left( e^{\imath2\pi \alpha q} \sum_{n=0}^{l_i-1} e^{-\imath\frac{2\pi}{N} \phi  n} + \sum_{n=l_i}^{N-1} e^{-\imath\frac{2\pi}{N} \phi  n} \right),
\end{align}
where \(\phi  = \alpha(p-q)+\nu_i+2Nc_1l_i.\) 
In addition, the \((p,q)\) entry of \(\mathbf{H}_i\) in \eqref{eq:H_i} can be written as
\begin{align}\label{eq:H_i_final}
H_i[p,q] ={\Lambda}_{c_2}[p,p]\, {W}[p,q]\, {\Lambda}_{c_2}^*[q,q] = e^{\imath2\pi c_2\left(q^2 - p^2\right)}\, {W}[p,q].
\end{align}
By substituting \eqref{eq:W_result} and \eqref{eq:I_sum_simplified} into \eqref{eq:H_i_final}, we obtain the closed-form expressions in \eqref{eq:SubHeff_Hi} - \eqref{SubHeff_Hi_Part2}.

\section*{ {Appendix B\\Derivation of the Transmitted Signal in \eqref{eq:NonAFDM3}}}
 {The time-domain signal in \eqref{eq:NonAFDM2} can be expressed in matrix form as  
\begin{align}\label{eq:s1}
    \mathbf{s}' = \bar{\boldsymbol{\Lambda}}_{c_1}^H \, \bar{\mathbf{F}}^H \, \bar{\boldsymbol{\Lambda}}_{c_2}^H \, \mathbf{x}',
\end{align}
where \(\mathbf{x}' = \begin{bmatrix} \mathbf{x}^T,& \mathbf{0}_{(N'-N)\times1}^T \end{bmatrix}^T\in \mathbb{C}^{N'\times 1}\) denotes the zero-padded symbol vector, and the extended diagonal matrix \(\bar{\boldsymbol{\Lambda}}_{c}\) is defined as
\begin{align}\label{eq:s1_1}
\bar{\boldsymbol{\Lambda}}_{c} =
\begin{bmatrix}
\boldsymbol{\Lambda}_{c} & \mathbf{0}_{N\times(N'-N)} \\
\mathbf{0}_{(N'-N)\times N} & \mathbf{B}_{c}
\end{bmatrix},
\end{align}
with \(\mathbf{B}_{c}=\mathrm{diag}\left(e^{-\imath 2\pi c n^2}, \, n = N, \ldots, N'-1 \right)\). 
The $N'$-point DFT matrix $\bar{\mathbf{F}}\in \mathbb{C}^{N'\times N'}$ is given by
$\bar{\mathbf{F}}[m,n] = \frac{1}{\sqrt{N'}} e^{-\imath 2\pi \frac{mn}{N'}},\ 0\le m,n\le N'-1$.
Since $N'=N/\alpha$, we have
$\frac{1}{\sqrt{N'}} e^{-\imath 2\pi \frac{mn}{N'}}
=\sqrt{\alpha}\cdot \frac{1}{\sqrt{N}} e^{-\imath 2\pi \frac{\alpha mn}{N}}
\triangleq \sqrt{\alpha}\,\mathbf{F}_{\alpha}[m,n]$ for $0\le m,n\le N-1$. Therefore, $\bar{\mathbf{F}}$
can be partitioned as
\begin{align}\label{eq:DFT}
\bar{\mathbf{F}} =
\begin{bmatrix}
\sqrt{\alpha}\,\mathbf{F}_{\alpha} & \mathbf{E} \\
\mathbf{G} & \mathbf{J}
\end{bmatrix},
\end{align}
where $\mathbf{E}\in\mathbb{C}^{N\times (N'-N)}$, $\mathbf{G}\in\mathbb{C}^{(N'-N)\times N}$, and
$\mathbf{J}\in\mathbb{C}^{(N'-N)\times (N'-N)}$
represent the corresponding subblocks of \(\bar{\mathbf{F}}\). Substituting \eqref{eq:s1_1} and \eqref{eq:DFT} into \eqref{eq:s1}, we obtain
\begin{align}\label{eq:s_new}
\mathbf{s}' &=
\begin{bmatrix}
\sqrt{\alpha}\, \boldsymbol{\Lambda}_{c_1}^{H} \mathbf{F}_{\alpha}^{H} \boldsymbol{\Lambda}_{c_2}^{H} \mathbf{x} \\
\mathbf{B}_{c_1}^H \mathbf{E}^H \boldsymbol{\Lambda}_{c_2}^H \mathbf{x}
\end{bmatrix}
=
\begin{bmatrix}
\sqrt{\alpha}\,\mathbf{s}_{\alpha}\\
\mathbf{B}_{c_1}^H \mathbf{E}^H \boldsymbol{\Lambda}_{c_2}^H \mathbf{x}
\end{bmatrix}.
\end{align}
This confirms that \(\mathbf{s}_{\alpha}\) in \eqref{eq:NonAFDM3} corresponds to the first \(N\) samples of \(\mathbf{s}'\) after scaling by \(1/\sqrt{\alpha}\), thus completing the derivation.}


\end{document}